\documentclass[11pt,a4paper]{article}
\usepackage{jheppub}
\usepackage[T1]{fontenc} 
\usepackage{slashed}
\usepackage{amsmath}
\usepackage{verbatim}
\usepackage[colorinlistoftodos]{todonotes}
\usepackage{xcolor}

\title{NLO JIMWLK evolution with massive quarks}

\def\BenGurion{Department of Physics, Ben-Gurion University of the Negev, Beer-Sheva 84105, Israel}
\def\TUMPhys{Physik Department, Technische Universit\"at M\"unchen, James-Franck-Strasse 1, 85748 Garching, Germany}

\author[a,b]{Lin Dai}
\author[a]{and Michael Lublinsky}

\affiliation[a]{\BenGurion}
\affiliation[b]{\TUMPhys}

\emailAdd{lin.dai@tum.de}
\emailAdd{lublinm@bgu.ac.il}

\abstract{
NLO evolution of the Jalilian-Marian-Iancu-McLerran-Weigert-Leonidov-Kovner (JIMWLK) equation with massless quarks was derived a few years ago. We make a step further to compute the  evolution kernels focusing on the effects due to finite quark masses. To this goal, the light-cone wave function of a fast moving dilute hadronic projectile is computed up to ${\cal O}(g^3)$  in  QCD coupling constant.
Compared with the massless case, 
a new IR divergence emerges,
which is eventually canceled by a mass dependent  counter term.  Our results 
extend the theoretical tools used in physics of gluon saturation and aim at improving 
precision in future phenomenological applications.
}

\begin{document}

\maketitle 


\section{Introduction} 

High energy evolution of a hadronic observable $\cal O$ in a dilute-on-dense scattering (DIS-like processes) is governed by the  JIMWLK   equation of the form 
\begin{equation}
\dot{\cal O} \,=\,-\,H^{JIMWLK}\,{\cal O}\,,
\end{equation}
which is 
a non-linear functional generalisation of  the BFKL  equation 
\cite{BFKL,BFKL1}. 

The JIMWLK evolution
\cite{cgc,cgc1,Jalilian-Marian:1997ubg,Kovner:2000pt,Ferreiro:2001qy}, 
or mainly its large $N_c$/mean field version, the Balitsky-Kovchegov (BK)
equation \cite{BK,BK2,BK1},
have been instrumental  for phenomenology of high energy QCD. So far,
most of the phenomenology  have been done using the LO formulation. Yet, in order to 
disentangle the gluon saturation effects it is critical to improve precision of the theoretical calculations
by going beyond the LO. A lot of progress has been achieved in this direction 
over the last fifteen years or so. Particularly, NLO formulation of both the BK equation 
\cite{Balitsky:2007feb} 
and JIMWLK Hamiltonian became available
\cite{Kovner:2013ona,Lublinsky:2016meo}
starting the era of NLO phenomenology
\cite{lappi,Beuf:2020dxl}. 
It was discovered that at NLO, there emerge  conceptual problems related to
 large energy independent new logarithmic corrections, threatening stability of the $\alpha_s$ expansion.  A lot of efforts have been invested over the years attempting  to fix these problems 
\cite{Kutak,Motyka,Beuf,Iancu:2015vea,Iancu:2015joa,SabioVera:2005tiv,Ducloue:2019ezk,Ducloue:2019jmy}.
While overcoming them is critical in order to gain full control over the calculations and achieve
 the desired NLO accuracy, these research directions  are in a sense orthogonal to our  goals  
below and hence would not be reviewed in any detail here. 

The NLO JIMWLK Hamiltonian has been derived for massless quarks only in previous literature, which at NLO 
contribute as loop corrections. Yet, it is known that quark masses could be significant phenomenologically, especially at realistic (not asymptotically large) energies. Say, at HERA, at relatively moderate Bjorken $x\sim 10^{-2}-
10^{-3}$, the charm quarks contribute to $F_2$ about 10\%  \cite{Gotsman:2002yy}. 
Notice that this is precisely the kinematical regime of future EIC operation. 
Inclusion of quark mass effects into evolution is an important additional step towards improving 
precision of the phenomenological tools. By this publication we are filling this gap.  Our work is complimentary to the most recent one \cite{Beuf:2021qqa}, which computed the mass effects in the photon impact factors.

In the next Section, we review the basics of the JIMWLK, particularly the Light Cone Wave Function (LCWF) formalism, which is central to our approach. The derivation will closely follow Ref. \cite{Lublinsky:2016meo}. 
Section \ref{sec:nlo-jimwlk} presents computation of the 
LCWF with massive quarks and  derives the quark mass-modified kernels of the JIMWLK Hamiltonian.
The last Section (Section \ref{sec:conclusions}) provides some concluding remarks.
Several Appendices supplement our calculations.

\section{From LCWF to  JIMWLK -- a brief review.}
\label{sec:basics}
In this section,  we revisit the key elements  needed for the derivation of the NLO JIMWLK equation with massive quarks.  We use the light cone quantization and the conventions used here closely follow those 
of Ref.~\cite{Lublinsky:2016meo}. 

\subsection{Light-Cone Quantization}
The light-cone (LC) coordinates $x^{\pm} \equiv  (x^0 \pm x^3)$ and $\mathbf{x} \equiv (x^1, x^2)$, 
and $p \cdot x = (p^+x^-+p^-x^+)/2-\mathbf{p}\cdot \mathbf{x}$. Throughout the paper, the light-cone gauge $A^{a+} = 0$  is used. Here 
 $a$ stands for the adjoint color index. In the LC gauge, the independent degrees of freedom (d.o.f.)  are  two transverse gluon components denoted below as $A^a_i$ with $i =1,2$. For fermion fields, the independent dofs are projected out by the light-cone projection operator 
 $\Lambda_{+}$; 
 $\psi_{+} \equiv \Lambda_{+} \psi$, where $\Lambda_{+} \equiv 
\gamma^0 \gamma^+/2$. 
The complementary projection of the fermion fields, i.e., $\psi_{-} \equiv \Lambda_{-} \psi$ with $\Lambda_{-} \equiv \gamma^0 \gamma^-/2$, is related with $\psi_{+}$ through equation of motion.

In the LC gauge the LC Hamiltonian with massive quarks reads \cite{Brodsky:1997de}
\begin{equation}
	H_{0} \equiv \int d x_{+} d^{2} \mathbf{x}\left[\frac{1}{2}\left(\partial_{i} A_{j}^{a}\right)^{2}+ \psi_{+}^{\dagger} \frac{-\partial_{i} \partial_{i}+m^2}{i\partial^{+}} \psi_{+}\right],
	\label{eq:kinetic-terms}
\end{equation}
for the kinetic part and the interaction part is \cite{Brodsky:1997de}:
\begin{eqnarray}
		H_{i n t} &\equiv & \int d x_{+} d^{2} \mathbf{x}\left[-g f^{a b c} \partial_{i} A_{j}^{a} A_{i}^{b} A_{j}^{c} +\frac{g^{2}}{4} f^{a b c} f^{a d e} A_{i}^{b} A_{j}^{c} A_{i}^{d} A_{j}^{e}\right. \nonumber\\
		&&-g f^{a b c}\left(\partial_{i} A_{i}^{a}\right) \frac{1}{\partial^{+}}\left(A_{j}^{b} \partial^{+} A_{j}^{c}\right)+\frac{g^{2}}{2} f^{a b c} f^{a d e} \frac{1}{\partial^{+}}\left(A_{i}^{b} \partial^{+} A_{i}^{c}\right) \frac{1}{\partial^{+}}\left(A_{j}^{d} \partial^{+} A_{j}^{e}\right) \nonumber\\
		&&+2 g^{2} f^{a b c} \frac{1}{\partial^{+}}\left(A_{i}^{b} \partial^{+} A_{i}^{c}\right) \frac{1}{\partial^{+}}\left(\psi_{+}^{\dagger} t^{a} \psi_{+}\right)+2 g^{2} \frac{1}{\partial^{+}}\left(\psi_{+}^{\dagger} t^{a} \psi_{+}\right) \frac{1}{\partial^{+}}\left(\psi_{+}^{\dagger} t^{a} \psi_{+}\right) \nonumber\\
		&&-2 g\left(\partial_{i} A_{i}^{a}\right) \frac{1}{\partial^{+}}\left(\psi_{+}^{\dagger} t^{a} \psi_{+}\right)-g \psi_{+}^{\dagger} t^{a}\left(i\alpha_{i} \partial_{i}+ m \beta\right) \frac{1}{i\partial^{+}}\left(\alpha_{j} A_{j}^{a} \psi_{+}\right) \nonumber\\
		&&-g \psi_{+}^{\dagger} t^{a} \alpha_{i} A_{i}^{a} \frac{1}{i \partial^{+}}\left(i\alpha_{j} \partial_{j} + m\beta \right) \psi_{+} \left.-i g^{2} \psi_{+}^{\dagger} t^{a} t^{b} \alpha_{i} A_{i}^{a} \frac{1}{\partial^{+}}\left(\alpha_{j} A_{j}^{b} \psi_{+}\right)\right]
		\label{eq:LC-Hamiltonian}
\end{eqnarray}
where 
$\alpha^i \equiv \gamma^0 \gamma^i$ and $\beta \equiv \gamma^0$.  
The mass dependent terms in the LC Hamiltonian include the fermion kinetic term in Eq.~(\ref{eq:kinetic-terms}) and the three point (two fermions and one gluon) interaction terms in the last two lines of Eq.~(\ref{eq:LC-Hamiltonian}). Four-point interaction terms are mass independent. While dealing with massive quarks, it is convenient to operate with four component spinors for fermion fields (note that this convention is different from Ref.~\cite{Lublinsky:2016meo}). The final results are, of course, independent of the conventions. 

For 4-component spinors, we use the following notation
\begin{equation}
	\chi_{\frac{1}{2}} \equiv \frac{1}{\sqrt{2}}\left(\begin{array}{l}
		1 \\
		0 \\
		1 \\
		0
	\end{array}\right), \quad
	\chi_{-\frac{1}{2}} \equiv \frac{1}{\sqrt{2}}\left(\begin{array}{l}
	0 \\
	1 \\
	0 \\
	-1
\end{array}\right),
\end{equation}
which satisfy
\begin{eqnarray}
	\chi_{\lambda_{1}}^{\dagger} \chi_{\lambda_{1}}=\delta_{\lambda_{1} \lambda_{2}}, \quad \chi_{\lambda_{1}}^{\dagger} \alpha^{3} \chi_{\lambda_{2}}=2 \lambda_{1} \delta_{\lambda_{1} \lambda_{2}}, \quad \sum_{\lambda} \chi_{\lambda} \chi_{\lambda}^{\dagger}=\Lambda_{+}.
\end{eqnarray}
In the Dirac representation, the 4-component fermion spinors have the following simple projections:
\begin{eqnarray}
	\Lambda_{+}u_{\pm \frac{1}{2}}(p) &= \sqrt{p^+}\chi_{\pm\frac{1}{2}}, \quad \Lambda_{+}v_{\pm \frac{1}{2}}(p) &= \sqrt{p^+}\chi_{\pm\frac{1}{2}}. 
	\label{eq:spinor-projection} 
\end{eqnarray}
Quantization of the fermion fields proceeds as usual by introducing the Fourier mode expansion
\begin{equation}
	\psi_{+}^\alpha(x)=\sum_{\lambda=\pm \frac{1}{2}} \chi_{\lambda} \int_{0}^{\infty} \frac{d k^{+}}{2 \pi} \int \frac{d^{2} \mathbf{k}}{(2 \pi)^{2}} \frac{1}{\sqrt{2}}\bigg(b^\alpha_{\lambda}\left(k^{+}, \mathbf{k}\right) e^{-i k \cdot x}+b_{\lambda}^{\alpha\dagger}\left(k^{+}, \mathbf{k}\right) e^{i k \cdot x}\bigg)
\end{equation}
where $\alpha$ is the fermion color index, the $1/\sqrt{2}$ factor comes from (using Eq.~(\ref{eq:spinor-projection}))
\begin{equation}
\frac{1}{\sqrt{2 k^{+}}} \Lambda_{+} u_{\lambda}\left(k^{+}, \mathbf{k}\right)=\frac{1}{\sqrt{2}} \chi_{\lambda}, \quad \frac{1}{\sqrt{2 k^{+}}} \Lambda_{+} v_{\lambda}\left(k^{+}, \mathbf{k}\right)=\frac{1}{\sqrt{2}} \chi_{\lambda}.
\end{equation}
Creation and annihilation operators satisfy the anti-commutation relations
\begin{equation}
\begin{aligned}
	\left\{b_{\lambda_{1}}^{\alpha}\left(k^{+}, \mathbf{k}\right), b_{\lambda_{2}}^{\beta \dagger}\left(p^{+}, \mathbf{p}\right)\right\} &=\left\{d_{\lambda_{1}}^{\alpha}\left(k^{+}, \mathbf{k}\right), d_{\lambda_{2}}^{\beta \dagger}\left(p^{+}, \mathbf{p}\right)\right\} \\
	&=(2 \pi)^{3} \delta_{\lambda_{1} \lambda_{2}} \delta^{\alpha \beta} \delta^{(2)}(\mathbf{k}-\mathbf{p}) \delta\left(k^{+}-p^{+}\right)
\end{aligned}
\end{equation} 
with all other anti-commutation relations vanishing.

The gluon fields Fourier expand as
\begin{equation}
	A_{i}^{a}(x)=\int_{0}^{\infty} \frac{d k^{+}}{2 \pi} \int \frac{d^{2} \mathbf{k}}{(2 \pi)^{2}} \frac{1}{\sqrt{2 k^{+}}}\bigg(a_{i}^{a}\left(k^{+}, \mathbf{k}\right) e^{-i k \cdot x}+a_{i}^{a \dagger}\left(k^{+}, \mathbf{k}\right) e^{i k \cdot x}\bigg).
\end{equation}
The commutation relations for creation and annihilation operators 
\begin{eqnarray}
	\left[a_{i}^{a}\left(k^{+}, \mathbf{k}\right), a_{j}^{b \dagger}\left(p^{+}, \mathbf{p}\right)\right]&=&(2 \pi)^{3} \delta^{a b} \delta_{i j} \delta\left(k^{+}-p^{+}\right) \delta^{(2)}(\mathbf{k}-\mathbf{p}) \nonumber \\
	\left[a_{i}^{a\dagger}\left(k^{+}, \mathbf{k}\right), a_{j}^{b \dagger}\left(p^{+}, \mathbf{p}\right)\right]&=&\left[a_{i}^{a}\left(k^{+}, \mathbf{k}\right), a_{j}^{b}\left(p^{+}, \mathbf{p}\right)\right] =0
\end{eqnarray}
The following normalization conventions for  free (anti-)quark and gluon states are used
\begin{eqnarray}
	\left|g_i^a\left(k^{+}, \mathbf{k}\right)\right\rangle&=&\frac{1}{(2 \pi)^{3 / 2}} a_i^{a\dagger}(k^+, \mathbf{k})|0\rangle,  \\
	\left|b_\lambda^\alpha \left(k^{+}, \mathbf{k}\right)\right\rangle&=&\frac{1}{(2 \pi)^{3 / 2}} b_\lambda^{\alpha\dagger}(k^+, \mathbf{k})|0\rangle, \quad \left|d_\lambda^\alpha \left(k^{+}, \mathbf{k}\right)\right\rangle=\frac{1}{(2 \pi)^{3 / 2}} d_\lambda^{\alpha\dagger}(k^+, \mathbf{k})|0\rangle .
\end{eqnarray}
Conventions for Fourier transformations 
\begin{eqnarray}
a_{i}^{a}\left(k^{+}, \mathbf{k}\right)&=&\int d^{2} \mathbf{z} e^{-i \mathbf{k} \cdot \mathbf{z}} a_{i}^{a}\left(k^{+}, \mathbf{z}\right), \\ 
b_{\lambda}^{\alpha}\left(k^{+}, \mathbf{k}\right)&=&\int d^{2} \mathbf{z} e^{-i \mathbf{k} \cdot \mathbf{z}} b_{\lambda}^{\alpha}\left(k^{+}, \mathbf{z}\right),  \ \ \ \ \ \ \ \ 
d_{\lambda}^{\alpha}\left(k^{+}, \mathbf{k}\right)=\int d^{2} \mathbf{z} e^{-i \mathbf{k} \cdot \mathbf{z}} d_{\lambda}^{\alpha}\left(k^{+}, \mathbf{z}\right), \ \ \ 
\end{eqnarray}
which also automatically fix the convention for  free states in (transverse) coordinate space, 
\begin{equation}
	\label{eq:convention-state-k-to-x}
	\left|g\left(k^{+}, \mathbf{k}\right)\right\rangle=\frac{1}{2 \pi} \int d^{2} \mathbf{z} e^{i \mathbf{k} \cdot \mathbf{z}}\left|g\left(k^{+}, \mathbf{z}\right)\right\rangle, ~~~~~~~etc.
\end{equation}

In this work, when loop correction are being computed, dimensional regularization with modified minimal subtraction scheme ($\overline{\rm MS}$ scheme) is used, i.e., 
\begin{equation}
	\int d^2 \mathbf{k} \cdots \quad \to \quad \mu^\epsilon \int d^{d} \mathbf{k} \cdots
\end{equation}
with $d \equiv 2-\epsilon$. Within the $\overline{\rm MS}$ scheme
\begin{equation}
	\mu^2 \to \frac{\bar{\mu}^2 e^{\gamma_{E}}}{4\pi}
\end{equation}
where $\bar{\mu}$ is short for the subtraction scale $\mu_{\overline{\rm MS}}$ in the $\overline{\rm MS}$ scheme, and $\gamma_{E}$ is the Euler-Mascheroni constant.

\subsection{LCWF of a fast moving projectile}
\label{subsec:wave-function-JIMWLK}
\begin{figure}
\centering
\includegraphics[width=12cm]{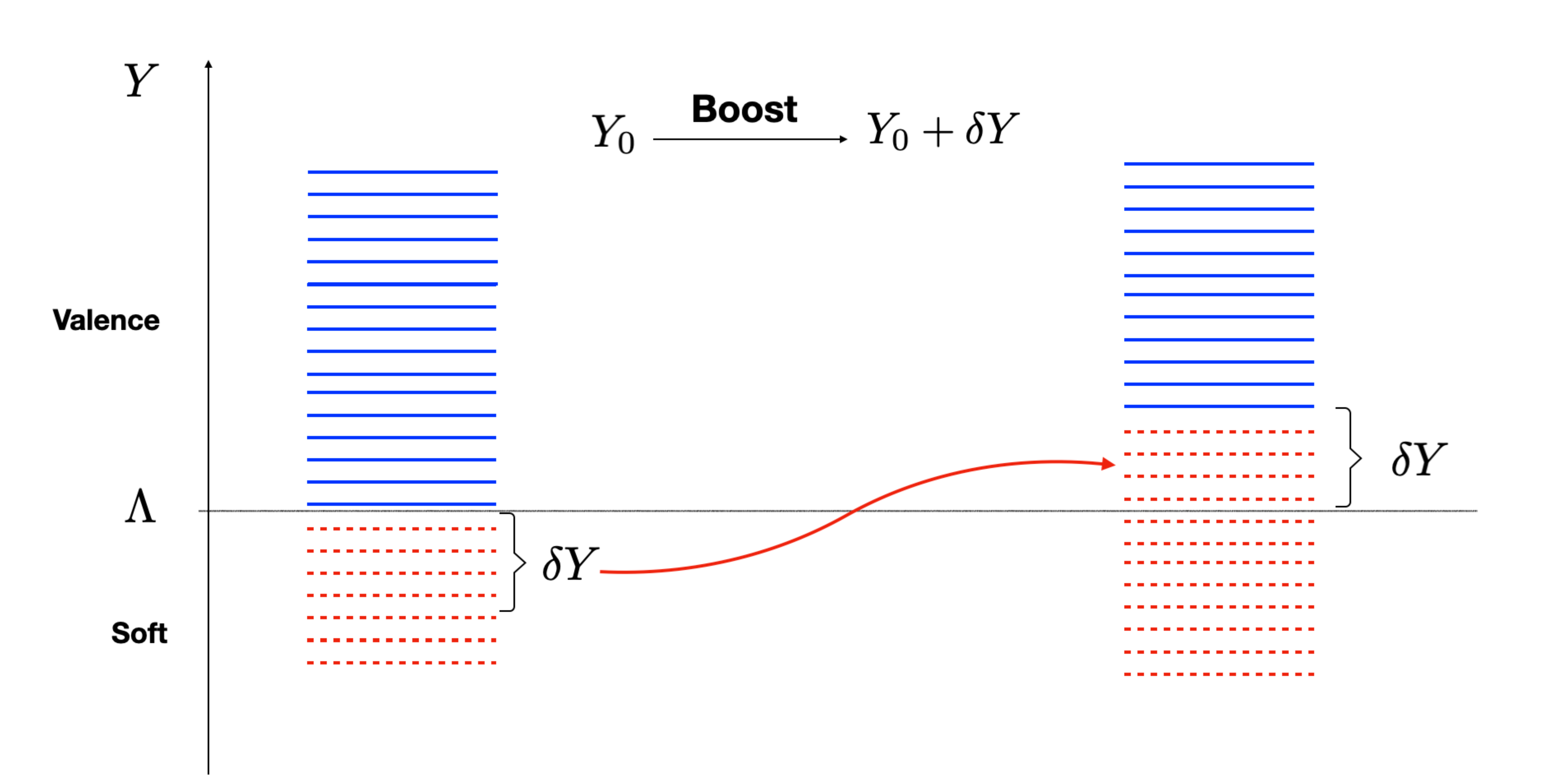}
\caption{The longitudinal momenta of the fast moving hadrons (projectiles) are divided into valence modes and soft modes with a separation scale $\Lambda$. 
After a differential boost of $\delta Y$, the soft modes within the range $\delta Y$ right below the cutoff $\Lambda$ will emerge  above $\Lambda$.}
\label{fig:boost-modes}
\end{figure}
LCWF approach provides a simple way of deriving the JIMWLK equation (see Ref.~\cite{Kovner:2005pe} for an introduction to the formalism). 
Following the Born-Oppenheimer approximation,
the longitudinal modes of a fast moving dilute hadron (projectile) are divided in longitudinal momentum 
into valence and soft modes, with a separation scale $\Lambda$, which implicitly depends on the collision target.
Below the scale $\Lambda$, the modes are too soft to contribute to the scattering amplitude. Energy of the collision
can be increased by boosting the projectile. How much a parton in a projectile is boosted is characterized by the boost parameter $Y$ (or rapidity), which is defined in the light-cone momentum as
\begin{equation}
k^+ = e^Y m_T, \quad \text{with} \quad m_T \equiv \sqrt{m^2+k_\perp^2},
\end{equation} 
where $m$ is the mass of the parton, and $k_\perp$ is the transverse momentum of the parton (or $\mathbf{k}$ in our convention setup above).
Since the rapidity is additive under a series of boosts, we have $k^+ \to k^{\prime +} = e^{\delta Y} k^+$ under a boost of $\delta Y$. After a boost of $\delta Y$, the soft modes within the range of $\delta Y$ right below $\Lambda$ emerge above $\Lambda$
and start to contribute to the scattering amplitude. 
This is the essence of the rapidity evolution of the LCWF. 

Assume that the projectile has already been boosted to a rapidity $Y_0$ 
with the quantum state described  by the valence modes only
\begin{equation}
	\left| \Psi \right\rangle_{Y_0} = |v\rangle.
\end{equation}
After the boost,  the soft modes emerge above $\Lambda$, resulting in a new LCWF of the projectile,  
\begin{equation}
	| \Psi \rangle_{Y_0+\delta Y} = | \psi \rangle \otimes | v \rangle
	\label{eq:boosted-state}
\end{equation}
where $| \psi \rangle$ refers to the soft modes of the state, whose longitudinal momenta lie in the window $(\Lambda, ~e^{\delta Y} \Lambda)$, as  shown in Fig.~\ref{fig:boost-modes}. Note that $\otimes$ here is symbolic since $| \psi \rangle$ depends on the valence state  $| v \rangle$ (soft and valence modes are entangled in a complex way).
Within the Born-Oppenheimer approximation, the valence part of the wave function in Eq.~(\ref{eq:boosted-state}) is assumed to be frozen while the soft part, $|\psi \rangle$, can be calculated perturbatively, order by order in QCD coupling $g$. The valence modes manifest themselves as a non-Abelian (non-commutative) background charge distribution, which  is intrinsically non-perturbative. 
The quark and gluon fields are decomposed into valence  and soft modes 
\begin{equation}
	\psi_{+}^{\alpha}(x)=\underline{\psi}_{+}^{\alpha}(x)+\overline{\psi}_{+}^{\alpha}(x), \quad A_{i}^{a}(x)=\underline{A}_{i}^{a}(x)+\overline{A}_{i}^{a}(x)
\end{equation}
where $\underline{\psi}$ and $\underline{A}$ are the soft modes, and $\overline{\psi}$ and $\overline{A}$ denote the valence modes. Substituting this decomposition into the  Hamiltonian ~(\ref{eq:LC-Hamiltonian}), one obtains an effective Hamiltonian for the soft-modes, including  their self-interaction, and interaction of the soft modes 
with the valence ~\cite{Lublinsky:2016meo}.  The latter are treated as  classical  background fields.

The effective soft Hamiltonian can be further simplified  due to eikonal approximation:
power suppressed terms (or higher order terms) will be ignored, where the power counting parameter will be a small parameter of the order $\lambda \sim p^+_{\rm soft}/p^+_{\rm valence}$. For instance, $\frac{1}{\partial^{+}}\left(\underline{A}_{j}^{b} \partial^{+} \overline{A}_{j}^{c}\right)$ is power suppressed compared to $\frac{1}{\partial^{+}}\left(\overline{A}_{j}^{b} \partial^{+} \overline{A}_{j}^{c}\right)$, where the former is counted as $\lambda^0$ and the latter as $\lambda^{-1}$. For the former, $\frac{1}{\partial^+}$ is counted as $1/p^+_{\rm valence}$ and $\partial^+ \overline{A}^c_j$ as $p^+_{\rm valence}$, so the term counts as $\lambda^0$. For the latter, $\frac{1}{\partial^+}$ is counted as $p^+_{\rm soft}$ (the sum of two valence momenta could be soft) and $\partial^+ \overline{A}^c_j$ as $p^+_{\rm valence}$, so the term is counted as $\lambda^{-1}$.

Below we will only focus on the terms in the effective Hamiltonian
which are relevant for the calculation of the quark mass effects.
These are
\begin{equation}\label{Hg}
	H_{g} \equiv-g \int d x_{+} d^{2} \mathbf{x}\left(\partial_{i} \underline{A}_{i}^{a}\right)\left(f^{a b c} \frac{1}{\partial^{+}}\left(\bar{A}_{j}^{b} \partial^{+} \bar{A}_{j}^{c}\right)+2 \frac{1}{\partial^{+}}\left(\bar{\psi}_{+}^{\dagger} t^{a} \bar{\psi}_{+}\right)\right),
\end{equation}
\begin{equation}\label{Hqqins}
	H_{q q-i n s t} \equiv 2 g^{2} \int d x_+ d^{2} \mathbf{x} \frac{1}{\partial^{+}}\left(\underline{\psi}_{+}^{\dagger} t^{a} \underline{\psi}_{+}\right)\left(f^{a b c} \frac{1}{\partial^{+}}\left(\bar{A}_{j}^{b} \partial^{+} \bar{A}_{j}^{c}\right)+2 \frac{1}{\partial^{+}}\left(\bar{\psi}_{+}^{\dagger} t^{a} \bar{\psi}_{+}\right)\right),
\end{equation}
which are in fact mass independent and identical to those used in Ref.~\cite{Lublinsky:2016meo}. The first term is
the leading soft gluon-valence interaction, responsible for the LO JIMWLK evolution. The second term is the instantaneous interaction of the quarks with the valence sector. It becomes important for quark bubble resummation.

The only quark mass dependent interaction is the gluon-quark-quark term
\begin{eqnarray}
\label{eq:gqq-Vertex}
H_{g q q} &\equiv&-g \int d x_{+} d^{2} \mathbf{x}\left(2\left(\partial_{i} \underline{A}_{i}^{a}\right) \frac{1}{\partial^{+}}\left(\underline{\psi}_{+}^{\dagger} t^{a} \underline{\psi}_{+}\right)+\underline{\psi}_{+}^{\dagger} t^{a}\left(i\alpha_{i} \partial_{i}+m\beta\right) \frac{1}{i\partial^{+}}\left(i\alpha_{j} \underline{A}_{j}^{a} \underline{\psi}_{+}\right)\right. \nonumber\\
	&&\left.+\underline{\psi}_{+}^{\dagger} t^{a}\left(i\alpha_{i} \underline{A}_{i}^{a}\right) \frac{1}{i\partial^{+}}\left(i\alpha_{j} \partial_{j}+m\beta \right)\underline{\psi}_{+}\right).
\end{eqnarray}
From Eq.~(\ref{eq:gqq-Vertex}), one obtains $H_{gqq}$
as a second quantized operator on the soft Hilbert space
\begin{equation}
	\begin{aligned}
		\mathrm{H}_{g q q} &=\sum_{\lambda_{1}, \lambda_{2}=\pm \frac{1}{2}} \int_{\Lambda}^{e^{\delta Y} \Lambda} \frac{d k^{+}}{2 \pi} \frac{d p^{+}}{2 \pi} d q^{+} \int \frac{d^{2} \mathbf{k}}{(2 \pi)^{2}} \frac{d^{2} \mathbf{p}}{(2 \pi)^{2}} d^{2} \mathbf{q} \frac{g t_{\alpha \beta}^{a}}{2 \sqrt{2 k^{+}}} \delta^{(3)}(k-p-q) \Gamma_{\lambda_{1} \lambda_{2}}^{i} \\
		& \times\left(a_{i}^{a}\left(k^{+}, \mathbf{k}\right) b_{\lambda_{1}}^{\alpha \dagger}\left(p^{+}, \mathbf{p}\right) d_{\lambda_{2}}^{\beta \dagger}\left(q^{+}, \mathbf{q}\right)+h . c .\right)
	\end{aligned}
\end{equation}
with
\begin{equation}
	\Gamma_{\lambda_{1} \lambda_{2}}^{i}=\chi_{\lambda_{1}}^{\dagger}\left[\frac{2 k_{i}}{k^{+}}-\frac{{\alpha} \cdot {\bf p}+m \beta}{p^{+}} \alpha_{i}-\alpha_{i} \frac{{\alpha} \cdot {\bf q}-m\beta}{q^{+}}\right] \chi_{\lambda_{2}}
\end{equation}
At NLO, a pair of soft quarks is emitted from a soft gluon with the mass dependent transition matrix element 
\begin{eqnarray}
	\left\langle\bar{q}_{\lambda_{2}}^{\delta}(q) q_{\lambda_{1}}^{\gamma}(p)\left|H_{g q q}\right| g_{i}^{a}(k)\right\rangle&=&\frac{g t_{\gamma \delta}^{a}}{8 \pi^{3 / 2} \sqrt{k^{+}}} \chi_{\lambda_{1}}^{\dagger}\left[\frac{2 \mathbf{k}^{i}}{k^{+}}-\frac{\alpha \cdot \mathbf{p}+m\beta}{p^{+}} \alpha^{i}-\alpha^{i} \frac{\alpha \cdot \mathbf{q}-m\beta}{q^{+}}\right] \chi_{\lambda_{2}}\nonumber \\ 
	&&\times~\delta^{(3)}(k-p-q)
\label{eq:Hgqq-ME}
\end{eqnarray}
Additional matrix elements, which are involved in the calculation below,  are all mass independent 
and are induced by the interactions (\ref{Hg}) and (\ref{Hqqins})  \cite{Lublinsky:2016meo}:
\begin{equation}
	\left\langle g_{i}^{a}(k)\left|H_{g}\right| 0\right\rangle=\frac{g \mathbf{k}^{i} \rho^{a}(-\mathbf{k})}{4 \pi^{3 / 2}\left|k^{+}\right|^{3 / 2}},
\label{eq:Hg-ME}
\end{equation}
and
\begin{equation}
	\left\langle\bar{q}_{\lambda_{2}}^{\beta}(q) q_{\lambda_{1}}^{\alpha}(p)\left|H_{q q\text{-inst}}\right| 0\right\rangle=\frac{g^{2} t_{\alpha \beta}^{a} \rho^{a}(-\mathbf{p}-\mathbf{q})}{8 \pi^{3}\left(p^{+}+q^{+}\right)^{2}} \chi_{\lambda_{1}}^{\dagger} \chi_{\lambda_{2}}.
	\label{eq:Hqq-inst-ME}
\end{equation}
Here $\rho^{a}(-\mathbf{p}) \equiv \rho_{g}^{a}(-\mathbf{p})+\rho_{q \bar{q}}^{a}(-\mathbf{p})$ with
\begin{equation}
	\rho_{g}^{a}(-\mathbf{p}) \equiv-i f^{a b c} \int_{\Lambda \delta Y}^{\infty} \frac{d k^{+}}{2 \pi} \int \frac{d^{2} \mathbf{k}}{(2 \pi)^{2}} a_{j}^{\dagger b}\left(k^{+}, \mathbf{k}\right) a_{j}^{c}\left(k^{+}, \mathbf{k}+\mathbf{p}\right)
\end{equation}
and
\begin{equation}
	\rho_{q \bar{q}}^{a}(-\mathbf{p}) \equiv t_{\alpha \beta}^{a} 
	\int_{e^{\delta Y} \Lambda}^{\infty} \frac{d k^{+}}{2 \pi} \int \frac{d^{2} \mathbf{k}}{(2 \pi)^{2}}\left(b_{\lambda}^{\alpha \dagger}\left(k^{+}, \mathbf{k}-\mathbf{p}\right) b_{\lambda}^{\beta}\left(k^{+}, \mathbf{k}\right)+d_{\lambda}^{\alpha}\left(k^{+}, \mathbf{k}\right) d_{\lambda}^{\beta \dagger}\left(k^{+}, \mathbf{k}-\mathbf{p}\right)\right)
\end{equation}
 Fourier transformed $\rho(\mathbf{p})$ is 
\begin{equation}
	\rho^{a}(\mathbf{p})=\int d^{2} \mathbf{x} e^{i \mathbf{p} \cdot \mathbf{x}} \rho^{a}(\mathbf{x}).
	\label{eq:rho-fourier-conv}
\end{equation}


\subsection{Deriving the NLO JIMWLK Hamiltonian from the LCWF}
The JIMWLK Hamiltonian is derived from the diagonal matrix element of the second quantized $\hat S$-matrix 
operator in the LCWF,
\begin{equation}\label{Sigma}
	\Sigma \equiv \langle \psi| \hat{S} -1|\psi \rangle
\end{equation} 
From $\Sigma$ one reads the Hamiltonian:
\begin{equation}
	\Sigma=e^{-\delta Y H_{J I M W L K}}-1 =-\delta Y H_{J I M W L K}+\frac{1}{2} \delta Y^{2} H_{J I M W L K}^{2} + {\cal O}(\delta Y^3).
\end{equation}

At high energies,  eikonal  approximation for the scattering is appropriate:
  the effect of a dense target on the projectile is only to color rotate the partons in the projectile,
\begin{equation}
	\hat{S} b_{i}^{\alpha \dagger}\left(x^{+}, \mathbf{x}\right)|0\rangle=S^{\beta \alpha}(\mathbf{x}) b_{i}^{\beta \dagger}\left(x^{+}, \mathbf{x}\right)|0\rangle, \quad \hat{S} d_{i}^{\alpha \dagger}\left(x^{+}, \mathbf{x}\right)|0\rangle=S^{\alpha \beta}(\mathbf{x}) d_{i}^{\beta \dagger}\left(x^{+}, \mathbf{x}\right)|0\rangle
\end{equation}
and
\begin{equation}
	\hat{S} a_{i}^{a \dagger}\left(x^{+}, \mathbf{x}\right)|0\rangle=S_{A}^{a b}(\mathbf{x}) a_{i}^{b \dagger}\left(x^{+}, \mathbf{x}\right)|0\rangle.
\end{equation}
Here $S^{\alpha \beta}$ is a Wilson line in the fundamental representation of the gauge fields along the target 
 light-cone direction, while $S^{ab}_A$ is the one in the adjoint representation. The action of $\hat{S}$ on the valence charge distribution $\rho^a(x)$ can be expressed succinctly as Lie derivatives with respect to the Wilson lines \cite{Kovner:2005jc}
\begin{equation}
	\rho_{g}^{a}(\mathbf{x}) \hat{S}|v\rangle=J_{R, a d j}^{a}(\mathbf{x})|\hat{S} v\rangle, \quad \hat{S} \rho_{g}^{a}(\mathbf{x})|v\rangle=J_{L, a d j}^{a}(\mathbf{x})|\hat{S} v\rangle,
\end{equation} 
with
\begin{equation}
	J_{R, a d j}^{a}(\mathbf{x}) \equiv-t r\left[S_{A}(\mathbf{x}) T^{a} \frac{\delta}{\delta S_{A}^{\dagger}(\mathbf{x})}\right],\quad J_{L, a d j}^{a}(\mathbf{x}) \equiv -t r\left[T^{a} S_{A}(\mathbf{x}) \frac{\delta}{\delta S_{A}^{\dagger}(\mathbf{x})}\right]
\end{equation}
and for quarks,
\begin{equation}
	\rho_{q \bar{q}}^{a}(\mathbf{x}) \hat{S}|v\rangle=J_{R, F}^{a}(\mathbf{x})|\hat{S} v\rangle, \quad \quad \quad \hat{S} \rho_{q \bar{q}}^a(\mathbf{x})|v\rangle=J_{L, F}^{a}(\mathbf{x})|\hat{S} v\rangle
\end{equation}
with
\begin{eqnarray}
	\begin{aligned}
		&J_{R, F}^{a}(\mathbf{x}) \equiv \operatorname{tr}\left[\frac{\delta}{\delta S^{T}(\mathbf{x})} S(\mathbf{x}) t^{a}\right]-\operatorname{tr}\left[\frac{\delta}{\delta S^{*}(\mathbf{x})} t^{a} S^{\dagger}(\mathbf{x})\right], \\
		&J_{L, F}^{a}(\mathbf{x}) \equiv \operatorname{tr}\left[\frac{\delta}{\delta S^{T}(\mathbf{x})} t^{a} S(\mathbf{x})\right]-\operatorname{tr}\left[\frac{\delta}{\delta S^{*}(\mathbf{x})} S^{\dagger}(\mathbf{x}) t^{a}\right].
	\end{aligned}
\end{eqnarray}
The Wilson lines $S$ parametrize the target. No additional information about the target or any explicit form of $S$ is needed for the derivation of the Hamiltonian.

The algorithm for the derivation is as follows. One first computes the LCWF $|\psi\rangle$ in perturbation theory. 
At LO, it is entirely determined by the matrix element (\ref{eq:Hg-ME}). The LCWF at LO has, in addition to the vacuum, a soft gluon component,
 \begin{eqnarray}
|\psi^{\rm LO}_g\rangle &=& -\int_\Lambda^{e^{\delta Y}\Lambda} dk^+\int d^2\mathbf{k} |g_i^a(k)\rangle \frac{\langle g_i^a(k)|H_g|0\rangle}{E_g(k)} \nonumber \\
&=& -\int_{\Lambda}^{e^{\delta \curlyvee} \Lambda} d k^{+} \int d^{2} \mathbf{k} \frac{g \mathbf{k}^{i}}{2 \pi^{3 / 2} \sqrt{k^{+}} \mathbf{k}^{2}} \rho^{a}(-\mathbf{k})\left|g_{i}^{a}(k)\right\rangle \nonumber \\
&=&  \int_{\Lambda}^{e^{\delta \curlyvee} \Lambda} \frac{d k^{+}}{\sqrt{k^{+}}} \int_{\mathbf{x}, \mathbf{z}} \frac{i g \ (\mathbf x- \mathbf z)^{i}}{2 \pi^{3 / 2} (\mathbf x- \mathbf z)^{2}} \rho^{a}(\mathbf{x})\left|g_{i}^{a}\left(k^{+}, \mathbf{z}\right)\right\rangle.
\label{eq:psig-LO}
 \end{eqnarray}
where $E_g(k) \equiv \frac{\mathbf{k}^2}{2k^+}$ is the gluon LC energy.

At a next step, one calculates $\Sigma$ in Eq.~(\ref{Sigma}), from which the JIMWLK Hamiltonian is read off as explained above. For massless quarks, 
the complete and detailed derivation along this line can be found in Ref. \cite{Lublinsky:2016meo}. Below 
we quote the final result for the sake of completeness of our presentation and easy comparison. 
In the next section, we will revise the calculation of \cite{Lublinsky:2016meo}  by extending it to include finite
quark masses.  

The NLO JIMWLK Hamiltonian with massless quarks are \cite{Kovner:2013ona, Lublinsky:2016meo}:
\begin{eqnarray}
		&&H_{J I M W L K}^{N L O}=\int_{\mathbf{x}, \mathbf{y}, \mathbf{z}} K_{J S J}(\mathbf{x}, \mathbf{y}, \mathbf{z})\left[J_{L}^{a}(\mathbf{x}) J_{L}^{a}(\mathbf{y})+J_{R}^{a}(\mathbf{x}) J_{R}^{a}(\mathbf{y})-2 J_{L}^{a}(\mathbf{x}) S_{A}^{a b}(\mathbf{z}) J_{R}^{b}(\mathbf{y})\right] \nonumber\\
		&&+\int_{\mathbf{x}, \mathbf{y}, \mathbf{z}, \mathbf{z}^{\prime}} K_{J S S J}\left(\mathbf{x}, \mathbf{y}, \mathbf{z}, \mathbf{z}^{\prime}\right)\left[f^{a b c} f^{d e f} J_{L}^{a}(\mathbf{x}) S_{A}^{b e}(\mathbf{z}) S_{A}^{c f}\left(\mathbf{z}^{\prime}\right) J_{R}^{d}(\mathbf{y})-N_{c} J_{L}^{a}(\mathbf{x}) S_{A}^{a b}(\mathbf{z}) J_{R}^{b}(\mathbf{y})\right] \nonumber\\
		&&+\int_{\mathbf{x}, \mathbf{y}, \mathbf{z}, \mathbf{z}^{\prime}} K_{q \bar{q}}\left(\mathbf{x}, \mathbf{y}, \mathbf{z}, \mathbf{z}^{\prime}\right)\left[2 J_{L}^{a}(\mathbf{x}) \operatorname{tr}\left[S^{\dagger}(\mathbf{z}) t^{a} S\left(\mathbf{z}^{\prime}\right) t^{b}\right] J_{R}^{b}(\mathbf{y})-J_{L}^{a}(\mathbf{x}) S_{A}^{a b}(\mathbf{z}) J_{R}^{b}(\mathbf{y})\right] \nonumber\\
		&&+\int_{\mathbf{w}, \mathbf{x}, \mathbf{y}, \mathbf{z}, \mathbf{z}^{\prime}} K_{J J S S J}\left(\mathbf{w}, \mathbf{x}, \mathbf{y}, \mathbf{z}, \mathbf{z}^{\prime}\right) f^{a c b}\left[J_{L}^{d}(\mathbf{x}) J_{L}^{e}(\mathbf{y}) S_{A}^{d c}(\mathbf{z}) S_{A}^{e b}\left(\mathbf{z}^{\prime}\right) J_{R}^{a}(\mathbf{w})\right. \nonumber\\
		&&\left.-J_{L}^{a}(\mathbf{w}) S_{A}^{c d}(\mathbf{z}) S_{A}^{b e}\left(\mathbf{z}^{\prime}\right) J_{R}^{d}(\mathbf{x}) J_{R}^{e}(\mathbf{y})+\frac{1}{3}\left(J_{L}^{c}(\mathbf{x}) J_{L}^{b}(\mathbf{y}) J_{L}^{a}(\mathbf{w})-J_{R}^{c}(\mathbf{x}) J_{R}^{b}(\mathbf{y}) J_{R}^{a}(\mathbf{w})\right)\right] \nonumber\\
		&&+\int_{\mathbf{w}, \mathbf{x}, \mathbf{y}, \mathbf{z}} K_{J J S J}(\mathbf{w}, \mathbf{x}, \mathbf{y}, \mathbf{z}) f^{b d e}\left[J_{L}^{d}(\mathbf{x}) J_{L}^{e}(\mathbf{y}) S_{A}^{b a}(\mathbf{z}) J_{R}^{a}(\mathbf{w})\right. \nonumber\\
		&&\left.-J_{L}^{a}(\mathbf{w}) S_{A}^{a b}(\mathbf{z}) J_{R}^{d}(\mathbf{x}) J_{R}^{e}(\mathbf{y})+\frac{1}{3}\left(J_{L}^{d}(\mathbf{x}) J_{L}^{e}(\mathbf{y}) J_{L}^{b}(\mathbf{w})-J_{R}^{d}(\mathbf{x}) J_{R}^{e}(\mathbf{y}) J_{R}^{b}(\mathbf{w})\right)\right]
\label{eq:jimwlk-hamiltonian-massless}
\end{eqnarray}
with
\begin{eqnarray}
	\begin{array}{r}
		K_{J J S S J}\left(\mathbf{w}, \mathbf{x}, \mathbf{y}, \mathbf{z}, \mathbf{z}^{\prime}\right)=-\frac{i \alpha_{s}^{2}}{4 \pi^{4}}\left(\frac{\left(Y^{\prime}\right)^{j} X^{i}}{\left(Y^{\prime}\right)^{2} X^{2}}-\frac{Y^{i}\left(X^{\prime}\right)^{j}}{\left(X^{\prime}\right)^{2} Y^{2}}\right) \\
		\times\left(\frac{\delta^{i j}}{2 Z^{2}}-\frac{\left(W^{\prime}\right)^{j} Z^{i}}{\left(W^{\prime}\right)^{2} Z^{2}}+\frac{W^{i} Z^{j}}{W^{2} Z^{2}}-\frac{W^{i}\left(W^{\prime}\right)^{j}}{W^{2}\left(W^{\prime}\right)^{2}}\right) \ln \left(\frac{W^{2}}{\left(W^{\prime}\right)^{2}}\right),
	\end{array}
\end{eqnarray}
where
\begin{equation}
	X \equiv \mathbf{x}-\mathbf{z}, X^{\prime} \equiv \mathbf{x}-\mathbf{z}^{\prime}, Y \equiv \mathbf{y}-\mathbf{z}, Y^{\prime} \equiv \mathbf{y}-\mathbf{z}^{\prime}, W \equiv \mathbf{w}-\mathbf{z}, W^{\prime} \equiv \mathbf{w}-\mathbf{z}^{\prime}, Z \equiv \mathbf{z}-\mathbf{z}^{\prime},
	\label{eq:XYZ-convection}
\end{equation}
\begin{equation}
	K_{J J S J}(\mathbf{w}, \mathbf{x}, \mathbf{y}, \mathbf{z})=-\frac{i \alpha_{s}^{2}}{4 \pi^{3}}\left(\frac{X \cdot W}{X^{2} W^{2}}-\frac{Y \cdot W}{Y^{2} W^{2}}\right) \ln \left(\frac{Y^{2}}{(X-Y)^{2}}\right) \ln \left(\frac{X^{2}}{(X-Y)^{2}}\right),
\end{equation}
\begin{eqnarray}
		&&K_{J S S J}\left(\mathbf{x}, \mathbf{y}, \mathbf{z}, \mathbf{z}^{\prime}\right)\nonumber \\
		&&=\frac{\alpha_{s}^{2}}{16 \pi^{4}}\left[\frac{4}{Z^{4}}+\left\{2 \frac{X^{2}\left(Y^{\prime}\right)^{2}+\left(X^{\prime}\right)^{2} Y^{2}-4(X-Y)^{2} Z^{2}}{Z^{4}\left(X^{2}\left(Y^{\prime}\right)^{2}-\left(X^{\prime}\right)^{2} Y^{2}\right)}+\frac{(X-Y)^{4}}{X^{2}\left(Y^{\prime}\right)^{2}-\left(X^{\prime}\right)^{2} Y^{2}}\right.\right.\nonumber \\
		&&\left.\times\left(\frac{1}{X^{2}\left(Y^{\prime}\right)^{2}}+\frac{1}{Y^{2}\left(X^{\prime}\right)^{2}}\right)+\frac{(X-Y)^{2}}{Z^{2}}\left(\frac{1}{X^{2}\left(Y^{\prime}\right)^{2}}-\frac{1}{Y^{2}\left(X^{\prime}\right)^{2}}\right)\right\} \ln \left(\frac{X^{2}\left(Y^{\prime}\right)^{2}}{\left(X^{\prime}\right)^{2} Y^{2}}\right)\nonumber \\
		&&\left.-\frac{2 I\left(\mathbf{x}, \mathbf{z}, \mathbf{z}^{\prime}\right)}{Z^{2}}-\frac{2 I\left(\mathbf{y}, \mathbf{z}, \mathbf{z}^{\prime}\right)}{Z^{2}}\right]+\widetilde{K}\left(\mathbf{x}, \mathbf{y}, \mathbf{z}, \mathbf{z}^{\prime}\right),
\end{eqnarray}
where 
\begin{eqnarray}
		&&I\left(\mathbf{x}, \mathbf{z}, \mathbf{z}^{\prime}\right) \equiv \frac{1}{X^{2}-\left(X^{\prime}\right)^{2}}\left(\frac{X^{2}+\left(X^{\prime}\right)^{2}}{Z^{2}}-\frac{X \cdot X^{\prime}}{X^{2}}-\frac{X \cdot X^{\prime}}{\left(X^{\prime}\right)^{2}}-2\right) \ln \left(\frac{X^{2}}{\left(X^{\prime}\right)^{2}}\right) \nonumber\\
		&&=\frac{1}{X^{2}-\left(X^{\prime}\right)^{2}}\left(\frac{X^{2}+\left(X^{\prime}\right)^{2}}{Z^{2}}+\frac{Z^{2}-X^{2}}{2\left(X^{\prime}\right)^{2}}+\frac{Z^{2}-\left(X^{\prime}\right)^{2}}{2 X^{2}}-3\right) \ln \left(\frac{X^{2}}{\left(X^{\prime}\right)^{2}}\right), 
\end{eqnarray}
and
\begin{eqnarray}
		&&\widetilde{K}\left(\mathbf{x}, \mathbf{y}, \mathbf{z}, \mathbf{z}^{\prime}\right)=\frac{\alpha_{s}^{2}}{16 \pi^{4}}\left(\frac{\left(Y^{\prime}\right)^{2}}{\left(X^{\prime}\right)^{2} Z^{2} Y^{2}}-\frac{Y^{2}}{Z^{2} X^{2}\left(Y^{\prime}\right)^{2}}+\frac{1}{Z^{2}\left(Y^{\prime}\right)^{2}}-\frac{1}{Z^{2} Y^{2}}+\frac{(X-Y)^{2}}{X^{2} Z^{2} Y^{2}}\right. \nonumber\\
		&&\left.-\frac{(X-Y)^{2}}{\left(X^{\prime}\right)^{2} Z^{2}\left(Y^{\prime}\right)^{2}}+\frac{(X-Y)^{2}}{\left(X^{\prime}\right)^{2} X^{2}\left(Y^{\prime}\right)^{2}}-\frac{(X-Y)^{2}}{X^{2}\left(X^{\prime}\right)^{2} Y^{2}}\right) \ln \left(\frac{X^{2}}{\left(X^{\prime}\right)^{2}}\right)+(\mathbf{x} \leftrightarrow \mathbf{y}),
\end{eqnarray}
\begin{eqnarray}
		&&K_{J S J}(\mathbf{x}, \mathbf{y}, \mathbf{z})=-\frac{\alpha_{s}^{2}}{16 \pi^{3}}\left(\frac { ( X - Y ) ^ { 2 } } { X ^ { 2 } Y ^ { 2 } } \left[b \ln \left((X-Y)^{2} \mu_{\overline{MS}}^{2}\right)-b \frac{X^{2}-Y^{2}}{(X-Y)^{2}} \ln \left(\frac{X^{2}}{Y^{2}}\right)\right.\right. \nonumber \\
		&&\left.+2 b(\gamma_E-\ln 2)+\left(\frac{67}{9}-\frac{\pi^{2}}{3}\right) N_{c}-\frac{10}{9} N_{f}\right]-\left[\frac{1}{X^{2}}+\frac{1}{Y^{2}}\right]\left[2 b(\gamma_E-\ln 2) N_{c}\right. \nonumber \\
		&&\left.\left.+\left(\frac{67}{9}-\frac{\pi^{2}}{3}\right)-\frac{10}{9} N_{f}\right]-\frac{b}{X^{2}} \ln X^{2} \mu_{\overline{M S}}^{2}-\frac{b}{Y^{2}} \ln Y^{2} \mu_{\overline{M S}}^{2}\right)-\frac{N_{c}}{2} \int_{\mathbf{z}^{\prime}} \widetilde{K}\left(\mathbf{x}, \mathbf{y}, \mathbf{z}, \mathbf{z}^{\prime}\right),\nonumber\\
\end{eqnarray}
where $b$ is the coefficient of the leading order QCD $\beta$-function
\begin{equation}
b=\frac{11N_C-2N_f}{3}.
\end{equation}
\begin{eqnarray}
	K_{q \bar{q}}\left(\mathbf{x}, \mathbf{y}, \mathbf{z}, \mathbf{z}^{\prime}\right) &=&\frac{\alpha_{s}^{2} N_{f}}{8 \pi^{4}}\left(\frac{2}{Z^{4}}-\frac{\left(X^{\prime}\right)^{2} Y^{2}+\left(Y^{\prime}\right)^{2} X^{2}-(X-Y)^{2} Z^{2}}{Z^{4}\left(X^{2}\left(Y^{\prime}\right)^{2}-\left(X^{\prime}\right)^{2} Y^{2}\right)}\left(\frac{X^{2}\left(Y^{\prime}\right)^{2}}{\left(X^{\prime}\right)^{2} Y^{2}}\right)\right. \nonumber\\
	&&\left.-\frac{I_{f}\left(\mathbf{x}, \mathbf{z}, \mathbf{z}^{\prime}\right)}{Z^{2}}-\frac{I_{f}\left(\mathbf{y}, \mathbf{z}, \mathbf{z}^{\prime}\right)}{Z^{2}}\right),
\label{Kqq}
\end{eqnarray}
with
\begin{equation}
	I_{f}\left(\mathbf{x}, \mathbf{z}, \mathbf{z}^{\prime}\right) \equiv \frac{2}{Z^{2}}-\frac{2 X \cdot X^{\prime}}{Z^{2}\left(X^{2}-\left(X^{\prime}\right)^{2}\right)} \ln \left(\frac{X^{2}}{\left(X^{\prime}\right)^{2}}\right).
\end{equation}
Quark mass  corrections will affect, as will be shown in the next sections, the kernels $K_{q\bar{q}}$ and $K_{JSJ}$ only, which we will denote as $K_{q\bar{q}}^m$ and $K_{JSJ}^m$
(the upperscript $m$ indicates dependence on mass).
The other kernels have been presented for completeness only and are irrelevant for the rest of the paper.

\section{NLO JIMWLK with massive quarks}
\label{sec:nlo-jimwlk}

\subsection{From $\left| \psi_{q\bar{q}}\rangle \right.$ to $K^m_{q\bar{q} }$}

\begin{figure}[htb]
	\centering
	\includegraphics[width=15cm]{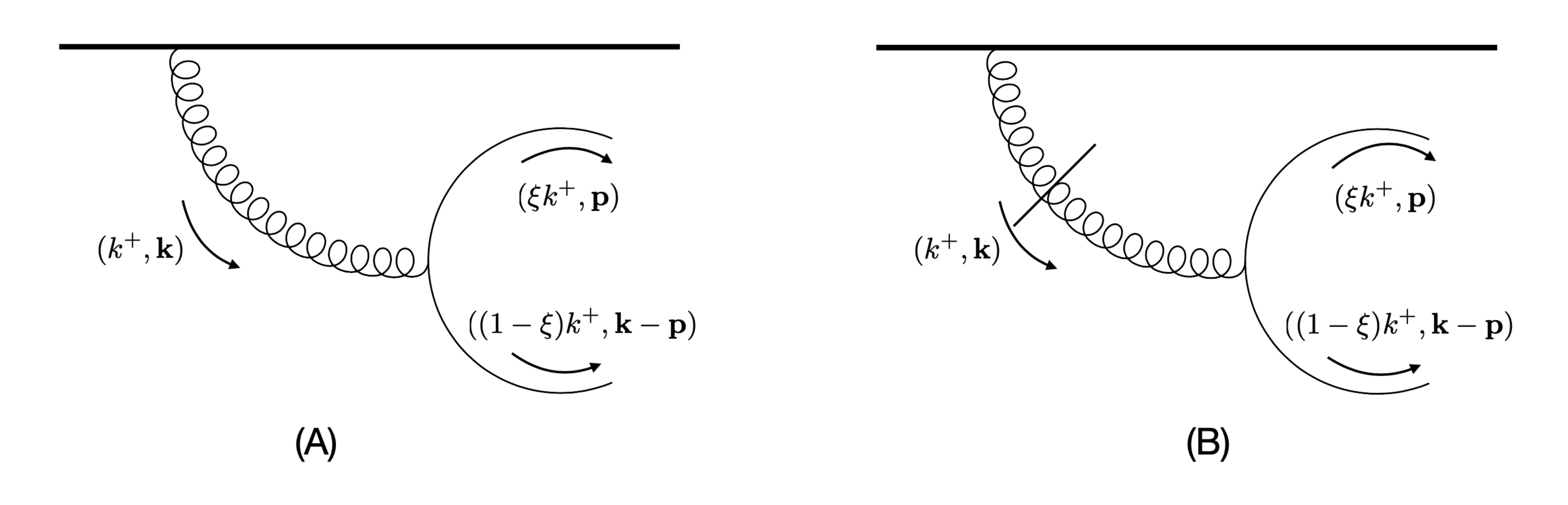}
	\caption{Quark emission diagrams, where $\xi$ is the defined as $\xi \equiv p^+/k^+$. The horizontal thick 
	 lines are valence charge current. Diagram (A) corresponds to the interaction $H_{gqq}$  (Eq.~\ref{eq:Hgqq-ME}), and (B)  to the instantaneous quark pair emission from the vertex $H_{qq\text{-inst}}$ (Eq.~\ref{eq:Hqq-inst-ME}). }
	\label{fig:qq-diagram}
\end{figure}
Similarly to the massless case \cite{Lublinsky:2016meo}, there are two terms in the LCWF with quark-antiquark state.
Those are  depicted in Fig.~\ref{fig:qq-diagram}. Let $|\psi_{q\bar{q}}^1\rangle$ 
denote the  wave function corresponding to Fig.~\ref{fig:qq-diagram} (A),
\begin{eqnarray}\label{pqq}
	\left|\psi_{q \bar{q}}^{1}\right\rangle & \equiv& \sum_{\lambda_{1}, \lambda_{2}} \int_{\Lambda}^{e^{\delta Y} \Lambda} d k^{+} d p^{+} d q^{+} \int d^{2} \mathbf{k} d^{2} \mathbf{p} d^{2} \mathbf{q} \nonumber\\
	&& \times\left|\bar{q}_{\lambda_{2}}^{\beta}(q) q_{\lambda_{1}}^{\alpha}(p)\right\rangle \frac{\left\langle\bar{q}_{\lambda_{2}}^{\beta}(q) q_{\lambda_{1}}^{\alpha}(p)\left|H_{g q q}\right| g_{i}^{a}(k)\right\rangle\left\langle g_{i}^{a}(k)\left|H_{g}\right| 0\right\rangle}{E_{q \bar{q}}(p, q) E_{g}(k)} 
\end{eqnarray}
The  energy denominator including  quark masses is
\begin{equation}
	E_{q \overline{q}}(p, q)=\frac{\mathbf{p}^{2}+m^{2}}{2 p^{+}}+\frac{\mathbf{q}^{2}+m^{2}}{2 q^{+}}
	\label{eq:energy-den}.
\end{equation}
The relevant matrix elements are given in Eq.~(\ref{eq:Hgqq-ME}) and (\ref{eq:Hg-ME}). Plugging in these elements, $|\psi_{q\bar{q}}^1\rangle$ becomes
\begin{eqnarray}
	\left|\psi_{q \bar{q}}^{1}\right\rangle & \equiv& \sum_{\lambda_{1}, \lambda_{2}} 
	\int_{\Lambda}^{e^{\delta Y} \Lambda} d k^{+} d p^{+} d q^{+} \int d^{2} \mathbf{k} d^{2} \mathbf{p} d^{2} \mathbf{q} \nonumber\\
	&& \times\left|\bar{q}_{\lambda_{2}}^{\beta}(q) q_{\lambda_{1}}^{\alpha}(p)\right\rangle \frac{g t_{\alpha \beta}^{a}}{8 \pi^{3 / 2} \sqrt{k^{+}}} \chi_{\lambda_{1}}^{\dagger}\left(\frac{2 k_{i}}{k^{+}}-\frac{\mathbf{\alpha} \cdot \mathbf{p}+\beta{m}}{p^{+}} \alpha_{i}-\alpha_{i} \frac{\mathbf{\alpha} \cdot \mathbf{q}-\beta{m}}{q^{+}}\right) \chi_{\lambda_{2}}\nonumber\\
	&& \times \delta^{3}(k-p-q) \frac{g k_{i} \rho(-\mathbf{k})}{4 \pi^{3 / 2}\left|k^{+}\right|^{3 / 2}} \frac{1}{\left(\frac{\mathbf{p}^{2}+m^{2}}{2 p^+}+\frac{\mathbf{q}^{2}+m^{2}}{2 q^{+}}\right) \frac{\mathbf{k}^{2}}{2 k^{+}}}.
	\label{eq:psi-qq-1}
\end{eqnarray}
Let $|\psi_{q\bar{q}}^2\rangle$ be the wave function component corresponding to the instantaneous quark pair emission diagram (B) in Fig.~\ref{fig:qq-diagram}, 
\begin{eqnarray}
	\left|\psi_{q \bar{q}}^{2}\right\rangle & \equiv& - \sum_{\lambda_{1}, \lambda_{2}} \int_{\Lambda}^{e^{\delta Y} \Lambda} d k^{+} d p^{+} \int d^{2} \mathbf{k} d^{2} \mathbf{p} \left|\bar{q}_{\lambda_{2}}^{\beta}(k-p) q_{\lambda_{1}}^{\alpha}(p)\right\rangle \frac{\left\langle\bar{q}_{\lambda_{2}}^{\beta}(k-p) q_{\lambda_{1}}^{\alpha}(p)\left|H_{qq\text{-inst}}\right| 0\right\rangle}{E_{q \bar{q}}(p, k-p)} \nonumber \\
	\label{eq:psi-qq-inst-pre}
\end{eqnarray}
where the minus sign in  front  is due to  perturbative expansion with one interaction insertion (see Eq.~(3.1) of \cite{Lublinsky:2016meo}, for example). Plugging in Eq.~(\ref{eq:Hqq-inst-ME}) and the massive quark energy denominator, we have
\begin{eqnarray}
	\left|\psi_{q \bar{q}}^{2}\right\rangle& =&
	 \int_{\Lambda}^{e^{\delta Y} \Lambda}d k^{+} \int_{0}^{k^{+}} d p^{+} \int d^{2}\mathbf{k} d^{2}\mathbf{p}\left|\bar{q}_{\lambda_{2}}^{\beta}(k-p) q_{\lambda_{1}}^{\alpha}(p)\right\rangle  \frac{g^{2} t_{\alpha \beta}^{a} \rho^{a}(-\mathbf{k}) \chi_{\lambda_{1}}^{\dagger} \chi_{\lambda_2}}{8 \pi^{3} k^{+2}\left(\frac{\mathbf{p}^{2}+m^{2}}{2 p^{+}}+\frac{(\mathbf{k}-\mathbf{p})^{2}+m^{2}}{2\left(k^{+}-p^{+}\right)}\right)} \nonumber\\
	\label{eq:psiqq-inst-2}
\end{eqnarray}
The sum of the two diagrams in Fig.~\ref{fig:qq-diagram} (A) and (B), 
$\left|\psi_{\bar{q}q \rho}\right\rangle=\left|\psi_{q \bar{q}}^{1}\right\rangle+\left|\psi_{q \bar{q}}^{2}\right\rangle$,
\begin{eqnarray}
	&&\left|\psi_{\bar{q}q \rho}\right\rangle =\nonumber \\
	 && \sum_{\lambda_{1} \lambda_{2}} \int_{\Lambda}^{e^{\delta Y} \Lambda} d k^{+} \int_{0}^{1} d \xi \int \frac{d^{2} \mathbf{k}}{(2 \pi)^{2}} \frac{d^{2}\mathbf{p}}{(2 \pi)^{2}}\frac{2 \pi g^{2} t_{\alpha \beta}^{a} \rho^{a} (-\mathbf{k}) \xi(1-\xi)}{(1-\xi) \mathbf{p}^{2}+\xi(\mathbf{k}-\mathbf{p})^{2}+m^{2}} \nonumber \\
	&& \quad \times \chi_{\lambda_{1}}^{\dagger}\left[\frac{k_{i}}{\mathbf{k}^{2}}\left(2 k_{i}-\frac{\mathbf{\alpha} \cdot \mathbf{p}+\beta m}{\xi} \alpha_{i}-\alpha_{i} \frac{{\alpha} \cdot(\mathbf{k}-\mathbf{p})-\beta{m}}{1-\xi}\right)-2\right] \chi_{\lambda_{2}} \left|\bar{q}_{\lambda_{2}}^{\beta}(k-p) q_{\lambda_{1}}^{\alpha}(p)\right\rangle \nonumber \\
	&&= - \sum_{\lambda_{1} \lambda_{2}} \int_{\Lambda}^{e^{\delta Y} \Lambda} d k^{+} \int_{0}^{1} d \xi \int \frac{d^{2} \mathbf{k}}{(2 \pi)^{2}} \frac{d^{2}\mathbf{p}}{(2 \pi)^{2}}\frac{2 \pi g^{2} t_{\alpha \beta}^{a} \rho^{a} (-\mathbf{k}) \xi(1-\xi)}{(1-\xi) \mathbf{p}^{2}+\xi(\mathbf{k}-\mathbf{p})^{2}+m^{2}} \nonumber \\
	&& \quad\times \chi_{\lambda_{1}}^{\dagger}\left[\frac{k_{i}}{\mathbf{k}^{2}}\left(\frac{\mathbf{\alpha} \cdot \mathbf{p}+\beta m}{\xi} \alpha_{i}+\alpha_{i} \frac{{\alpha} \cdot(\mathbf{k}-\mathbf{p})-\beta{m}}{1-\xi}\right)\right] \chi_{\lambda_{2}}\left|\bar{q}_{\lambda_{2}}^{\beta}(k-p) q_{\lambda_{1}}^{\alpha}(p)\right\rangle 
	\label{eq:psi-qq-sum}
\end{eqnarray} 
where $\xi \equiv p^+/k^+$. Notice that the instantaneous interaction  in Fig.~\ref{fig:qq-diagram} (B) is canceled exactly by one of the terms from diagram (A), which is shown in the second equality of Eq.~(\ref{eq:psi-qq-sum}).

With the wave function  Eq.~(\ref{eq:psi-qq-sum}) in hand, one can extract the kernel $K_{qq}^m$ of the 
JIMWLK Hamiltonian, as explained in the previous section.
 Denote the forward scattering amplitude for the quark anti-quark pair as 
\begin{equation}
	\Sigma_{\bar{q}q}^{\rm NLO} \equiv  \left\langle \psi_{\bar{q}q \rho} \right| \hat{S} \left|\psi_{\bar{q}q \rho}\right\rangle.
	\label{eq:sigma-qq-def}
\end{equation} 
Substituting Eq.~(\ref{eq:psi-qq-sum}) into Eq.~(\ref{eq:sigma-qq-def}) and using the  equality,
\begin{eqnarray}
	\left\langle \bar{q}_{\lambda_{4}}^{\delta}(u-v) q_{\lambda_3}^{\gamma}(v) \right| \hat{S} \left| \bar{q}_{\lambda_{2}}^{\beta}(k-p) q_{\lambda_{1}}^{\alpha}(p)\right\rangle &=& \frac{\delta_{\lambda_{1} \lambda_{3}} \delta_{\lambda_{2} \lambda_{4}} }{(2 \pi)^{4} k^{+}} \int_{\mathbf{z}, \mathbf{z}^{\prime}}e^{-i \mathbf{v} \cdot (\mathbf{z}-\mathbf{z}^{\prime})+i \mathbf{p} \cdot\left(\mathbf{z}-\mathbf{z}^{\prime}\right)-i \mathbf{u} \cdot \mathbf{z}^{\prime}+i \mathbf{k} \cdot \mathbf{z}^{\prime}}  \nonumber \\
	&& \times S^{\gamma \alpha}(\mathbf{z}) S^{\dagger\beta \delta}\left(\mathbf{z}^{\prime}\right) \delta\left(u^{+}-k^{+}\right) \delta(\xi-\theta).
	\label{eq:qqSqq}
\end{eqnarray}
Here again, $\xi \equiv p^+/k^+$ and $\theta \equiv v^+/u^+$.  After a few algebraic steps, 
detailed in Appendix~\ref{app:psi-qq}, we obtain
\begin{eqnarray}
	\Sigma_{\bar{q} q}^{\rm NLO}
	&=& \int_{\mathbf{x}, \mathbf{y}, \mathbf{z}, \mathbf{z}^{\prime}} \int_{\Lambda}^{e^{\delta Y}\Lambda} d k^{+} \int_{0}^{1} d \xi \int_{\mathbf{k}, \tilde{\mathbf{p}}, \mathbf{u}, \tilde{\mathbf{v}} } e^{-i \tilde{\mathbf{v}} \cdot {Z}+i\mathbf{u} \cdot\left({Y}^{\prime}-\xi {Z}\right)+i \tilde{\mathbf{p}} \cdot {Z}-i \mathbf{k} \cdot\left({X}^{\prime}-\xi {Z}\right)} \nonumber \\
	&& \times \frac{2 g^{4} J_{L}^{a}(\mathbf{x}) \operatorname{tr}\left[S(z) t^{a} S{\left(z^{\prime}\right)}^{\dagger} t^{b}\right] J_{R}^{b}(\mathbf{y})}{(2 \pi)^{10}\left(\xi(1-\xi) \mathbf{u}^{2}+\tilde{\mathbf{v}}^{2}+m^{2}\right)\left((1-\xi) \xi \mathbf{k}^{2}+{\tilde{\mathbf{p}}}+m^{2}\right)} \nonumber \\
	&& \times \Bigg[ 4 \xi^{2}(1-\xi)^{2}+2 \xi(1-\xi)(1-2 \xi)\left(\frac{\mathbf{k} \cdot \tilde{\mathbf{p}}}{\mathbf{k}^{2}}+\frac{\mathbf{u} \cdot \tilde{\mathbf{v}}}{\mathbf{u}^{2}}\right) \nonumber \\ 
	&& \quad \,\,\, +\frac{(1-2 \xi)^{2} \mathbf{k} \cdot \tilde{\mathbf{p}} \mathbf{u} \cdot \tilde{\mathbf{v}}+\mathbf{u} \cdot \mathbf{k} \tilde{\mathbf{v}} \cdot \tilde{\mathbf{p}}-\mathbf{u} \cdot \tilde{\mathbf{p}} \tilde{\mathbf{v}} \cdot \mathbf{k}+m^{2} \mathbf{u} \cdot \mathbf{k}}{\mathbf{u}^{2} \mathbf{k}^{2}} \Bigg].	
	\label{eq:sigma-qq-tilde}
\end{eqnarray} 
The multiple momentum integral $\int_{\mathbf{k}, \tilde{\mathbf{p}}, \mathbf{u}, \tilde{\mathbf{v}}}$ in Eq.~(\ref{eq:sigma-qq-tilde}) can be explicitly carried out with the help of the integrals collected in Appendix~\ref{app:integrals}, and the final result is
\begin{eqnarray}
	\Sigma_{\bar{q} q}^{\rm NLO}=&&-\delta Y ~\frac{g^{4}}{32\pi^6} \int_{\mathbf{x}, \mathbf{y}, \mathbf{z}, \mathbf{z}^{\prime}}J_{L}^{a}(\mathbf{x}) \operatorname{Tr}\left[S^{\dagger}(\mathbf{z}) t^{a} S\left(\mathbf{z}^{\prime}\right) t^{b}\right] J_{R}^{b}(\mathbf{y})\nonumber\\ 
	&& \int_{0}^{1} d \xi \left\{ \frac{1}{ Z^{4}\left((1-\xi)\left(X^{\prime}\right)^{2}+\xi X^{2}\right)\left((1-\xi)\left(Y^{\prime}\right)^{2}+\xi Y^{2}\right)} \left[-4 \xi^{2}(1-\xi)^{2} Z^{4}H_1\right. \right.\nonumber\\
	&&\left. \left.-2 \xi(1-\xi)(1-2 \xi)\left((X^{\prime} \cdot Z-\xi Z^2)H_2+(Y^{\prime} \cdot Z- \xi Z^{2})Z^{2} H_3\right) \right. \right. \nonumber\\
	&&\left. \left. +\left(4 \xi(1-\xi)\left(X^{\prime} \cdot Z-\xi Z^{2}\right)\left(Y^{\prime} \cdot Z-\xi Z^{2}\right)-\left(X^{\prime}-\xi Z\right) \cdot\left(Y^{\prime}-\xi Z\right) Z^{2}\right) H_4 \right] \right. \nonumber\\
	&&-4 m^{2} \frac{\left(X^{'}-\xi Z\right) \cdot\left(Y^{\prime}-\xi Z\right)}{\left(X^{\prime}-\xi Z\right)^{2}\left(Y^{\prime}-\xi Z\right)^{2}} H_{5}\Bigg\}
	\label{eq:sigmaqq}
\end{eqnarray}
where 
$H_1\equiv H_1(m\Xi,m\Upsilon)$ and $H_i\equiv H_i(m\Xi,m\Upsilon,mZ)$ for $i=2,...,5$ with
\begin{equation}
	\Xi \equiv \sqrt{\frac{X'^2}{\xi}+\frac{X^2}{1-\xi}}, ~ \Upsilon \equiv \sqrt{\frac{Y'^2}{\xi}+\frac{Y^2}{1-\xi}}.
\end{equation}
The functions $H_i$ are defined below in terms of the Modified Bessel functions of the second kind $K_0$ and $K_1$:
\begin{eqnarray}
	H_{1}(x, y)&=&x y K_{1}(x) K_{1}(y),\\
	H_{2}(x, y, z)&=&\frac{y K_{1}(y)}{x^{2}-z^{2}}\left[x^{2} z K_{1}(z)-z^{2} x K_{1}(x)\right],\\
	H_{3}(x, y, z)&=&H_{2}(y, x, z),\\
	H_{4}(x, y, z)&=&\frac{1}{x^{2}-z^{2}} \frac{1}{y^{2}-z^{2}}\left(x^{2} z K_{1}(z)-z^{2} x K_{1}(x)\right)\left(y^{2} z K_{1}(z)-z^{2} y K_{1}(y)\right),\\
	H_{5}(x, y, z)&=&\left(K_{0}(x)-K_{0}(z)\right)\left(K_{0}(y)-K_{0}(z)\right).
\end{eqnarray}
To recover the massless limit, $m\rightarrow 0$, we consider the asymptotic behavior
of these functions as $x\to 0$:
\begin{eqnarray}
	x K_1(x) &\to& 1, \text{ as } x\to 0,  \nonumber \\
	K_0(x) &\to& -\ln \frac{x}{2}-\gamma_E , \text{ as } x\to 0,  \nonumber \\
	H_1(mx,my) &\to& 1, \text{ as } m\to 0,  \nonumber \\
	H_i(mx,my,mz) &\to& 1, \text{ as } m\to 0, \text{ where } i=2,3,4, \nonumber \\
	H_5(mx,my,mz) &\to& \ln \frac{x}{z} \ln \frac{y}{z}, \text{ as } m\to 0.
\end{eqnarray}
With these asymptotic expressions, Eq.~(\ref{eq:sigmaqq}) reproduces the result for massless quarks, which has already been obtained in Ref.~\cite{Lublinsky:2016meo} (Eq.~(H.3) therein). 

Eq.~(\ref{eq:sigmaqq}) can be massaged a little bit further to take the following form
\begin{eqnarray}
	\Sigma_{q \bar{q}}^{\rm NLO}&=& -\delta Y ~\frac{g^{4}}{64\pi^6} \int_{\mathbf{x}, \mathbf{y}, \mathbf{z}, \mathbf{z}^{\prime}}J_{L}^{a}(\mathbf{x}) \operatorname{Tr}\left[S^{\dagger}(\mathbf{z}) t^{a} S\left(\mathbf{z}^{\prime}\right) t^{b}\right] J_{R}^{b}(\mathbf{y})\nonumber\\ 
	&& \int_{0}^{1} d \xi  \Bigg\{ \frac{1}{ Z^{4}\left((1-\xi)\left(X^{\prime}\right)^{2}+\xi X^{2}\right)\left((1-\xi)\left(Y^{\prime}\right)^{2}+\xi Y^{2}\right)}  \nonumber\\
	&& \Bigg[\xi(1-\xi)\left(X^{\prime 2}-X^{2}\right)\left(Y^{\prime 2}-Y\right) H_{4} + Z^2 \Bigg( 2Z^2 \Big(  2\xi(1-\xi)(1-2\xi+2\xi^2)  H_{4}	\nonumber\\
	&& -\xi(1-\xi)(1-2 \xi)^{2}\left(H_{2}+H_{3}\right)-4 \xi^{2}(1-\xi)^{2} H_{1}\Big)- X^{\prime 2}\Big( (-2\xi(1-\xi)(1-2 \xi) \nonumber\\
	&& +(1-\xi)) H_{4}+2\xi(1-\xi)(1-2 \xi) H_{2}\Big)- Y^{\prime 2}\Big( (-2\xi(1-\xi)(1-2 \xi)+(1-\xi)) H_{4} \nonumber\\
	&&+2\xi(1-\xi)(1-2 \xi) H_{3}\Big) - X^{2}\Big( (2\xi(1-\xi)(1-2 \xi)-\xi) H_{4}-2\xi(1-\xi)(1-2 \xi) H_{2}\Big) \nonumber\\
	&& - Y^{2}\Big( (2\xi(1-\xi)(1-2 \xi)-\xi) H_{4}-2\xi(1-\xi)(1-2 \xi) H_{3}\Big)-(X-Y)^2 H_4 \Bigg)\Bigg]\nonumber\\
	&&-4 m^{2} \frac{\left(X^{'}-\xi Z\right) \cdot\left(Y^{\prime}-\xi Z\right)}{\left(X^{\prime}-\xi Z\right)^{2}\left(Y^{\prime}-\xi Z\right)^{2}} H_{5}
	\Bigg\}	
	\label{eq:sigmaqqFinal}
\end{eqnarray}
From the expression above one reads off the quark mass-modified kernel $K_{q\bar{q}}^m$:
\begin{eqnarray}
	&&K_{q\bar{q}}^m(\mathbf{x,y,z,z'}) =  \frac{\alpha_s^2}{4\pi^4} \int_{0}^{1} d \xi  \Bigg\{ \frac{1}{ Z^{4}\left((1-\xi)\left(X^{\prime}\right)^{2}+\xi X^{2}\right)\left((1-\xi)\left(Y^{\prime}\right)^{2}+\xi Y^{2}\right)} \nonumber \\
	&& \times \Bigg[\xi(1-\xi)\left(X^{\prime 2}-X^{2}\right)\left(Y^{\prime 2}-Y\right) H_{4} \nonumber\\
	&& + Z^2 \Bigg( 2Z^2 \Big(  2\xi(1-\xi)(1-2\xi+2\xi^2)  H_{4}-\xi(1-\xi)(1-2 \xi)^{2}\left(H_{2}+H_{3}\right)-4 \xi^{2}(1-\xi)^{2} H_{1}\Big)	\nonumber\\
	&& - X^{\prime 2}\Big( (-2\xi(1-\xi)(1-2 \xi)+(1-\xi)) H_{4}+2\xi(1-\xi)(1-2 \xi) H_{2}\Big) \nonumber\\
	&& - Y^{\prime 2}\Big( (-2\xi(1-\xi)(1-2 \xi)+(1-\xi)) H_{4}+2\xi(1-\xi)(1-2 \xi) H_{3}\Big) \nonumber\\
	&& - X^{2}\Big( (2\xi(1-\xi)(1-2 \xi)-\xi) H_{4}-2\xi(1-\xi)(1-2 \xi) H_{2}\Big) \nonumber\\
	&& - Y^{2}\Big( (2\xi(1-\xi)(1-2 \xi)-\xi) H_{4}-2\xi(1-\xi)(1-2 \xi) H_{3}\Big)-(X-Y)^2 H_4 \Bigg)\Bigg]\nonumber\\
	&&-4 m^{2} \frac{\left(X^{'}-\xi Z\right) \cdot\left(Y^{\prime}-\xi Z\right)}{\left(X^{\prime}-\xi Z\right)^{2}\left(Y^{\prime}-\xi Z\right)^{2}} H_{5}
	\Bigg\}	.
	\label{eq:Kqq}
\end{eqnarray}
We have not succeeded to carry out the $\xi$-integral analytically. Hence, the final result cannot  be presented in terms of elementary functions.  Yet, the $\xi$-integral should be relatively easy for 
numerical evaluation.  The result  has been written for a single flavor only. Summation over 
different flavors/masses is obviously implied.

\subsection{ $K_{JSJ}^m$ with massive quarks}

There are two mass-dependent contributions to the kernel $K_{JSJ}^m$. The first one is direct contribution of the quark loop to the one gluon component of the LCWF
(see Fig.~\ref{fig:quark-loop-diagram}). Another one originates from the subtraction of the 
$K_{q\bar q}^m$ kernel, to be discussed below. 

\subsubsection{Quark loop effects in $| \psi_g \rangle$}

\begin{figure}[htb]
	\centering
	\includegraphics[width=11cm]{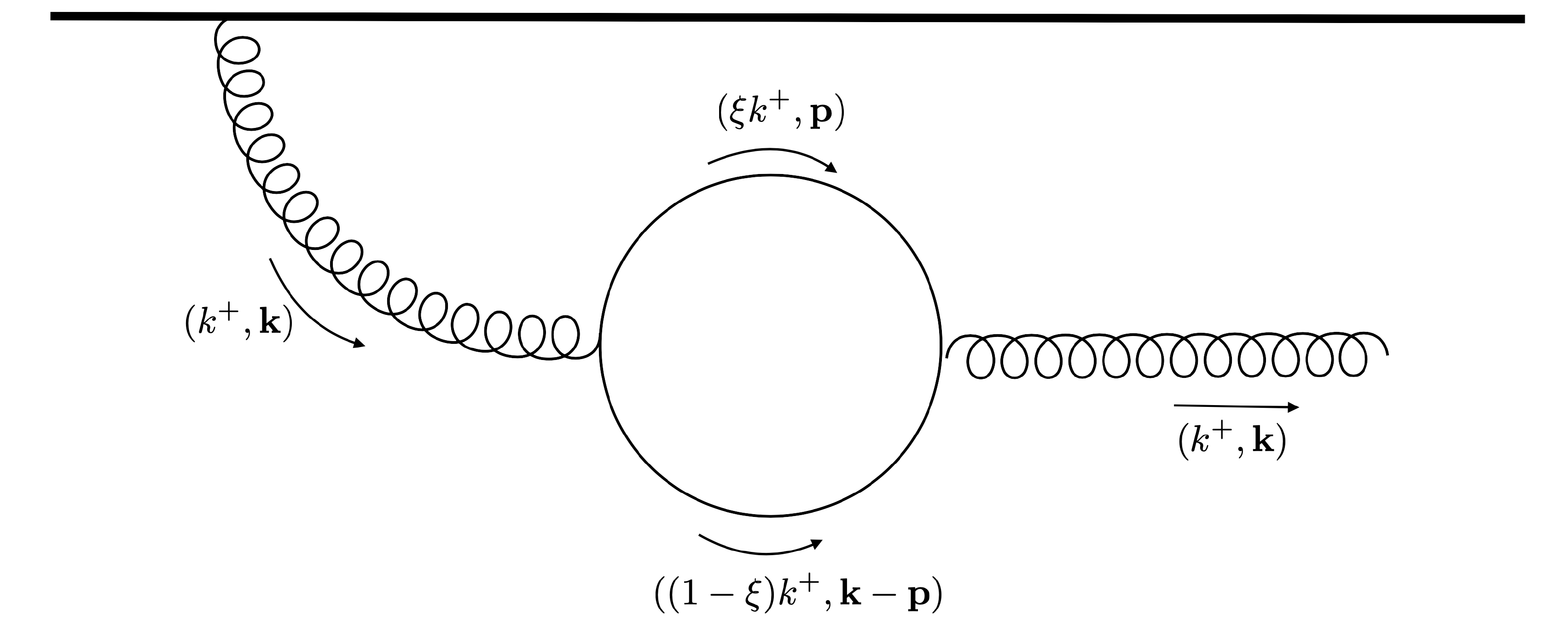}
	\caption{Quark loop correction for the single gluon emission diagram; $\xi$ is the defined as $\xi \equiv p^+/k^+$. The horizontal thick  line is the valence  current. }
	\label{fig:quark-loop-diagram}
\end{figure}
The quark loop correction to the single gluon  component of the LCWF is displayed 
in Fig.~{\ref{fig:quark-loop-diagram}}. In the massless limit, the diagram is UV divergent,  giving rise to running of 
the coupling (the $N_f$ term in the $\beta$-function), after standard  dimensional regularization is applied \cite{Lublinsky:2016meo}. 
While renormalization is straightforward in the massless case, the massive case turns out to be much more involved. 
Most of this subsection is devoted to carefully carrying out the renormalization procedure, which requires introduction of
a mass-dependent counter term.
 
By $| \psi_g^1 \rangle$ we denote the renormalized component. 
Un-renormalized component $|\psi_{g}^{1'}\rangle$  is expressed in perturbation theory,
\begin{eqnarray}
	\label{eq:psig-1}
	\left|\psi_{g}^{1'}\right\rangle &\equiv&-\int_{\Lambda}^{e^{\delta Y} \Lambda} d k^{+} d p^{+} d q^{+} d r^{+} \int d^{2} \mathbf{k} d^{2} \mathbf{p} d^{2} \mathbf{q} d^{2} \mathbf{r} \nonumber\\
	&& \times\left|g_{j}^{d}(r)\right\rangle \frac{\left\langle g_{j}^{d}(r)\left|H_{g q q}\right| q_{\lambda_{1}}^{\alpha}(p) \bar{q}_{\lambda_{2}}^{\beta}(q)\right\rangle\left\langle q_{\lambda_{1}}^{\alpha}(p) \bar{q}_{\lambda_{2}}^{\beta}(q)\left|H_{g q q}\right| g_{i}^{a}(k)\right\rangle\left\langle g_{i}^{a}(k)\left|H_{g}\right| 0\right\rangle}{E_{g}(r) E_{q \bar{q}}(p, q) E_{g}(k)}\nonumber\\
	\label{eq:psi-g1}
\end{eqnarray}
The  matrix element of $H_{gqq}$
is given in Eq.~(\ref{eq:Hgqq-ME}). The detailed calculation of Eq.~(\ref{eq:psig-1}) can be found 
in Appendix~\ref{app:psig-1}. The final, dimensionally regularized result  is
\begin{eqnarray}
\label{eq:psig-1-final}
&&|\psi_g^{1'} \rangle =\nonumber \\
&& -\mu^{\epsilon}\int_{\Lambda}^{e^{\delta Y}\Lambda} d k^{+} \int_{0}^{1} d \xi \int \frac{d^{2} \mathbf{k}}{ (2\pi)^{2}} \frac{d^{d} \tilde{\mathbf{p}}}{(2 \pi)^{d}}  \frac{g^{3} \rho^{a}(-\mathbf{k}) k_{i}\left((4\xi^2-4\xi+d){\tilde{\mathbf{p}}}^{2}+m^{2}d \right)}{2 \sqrt{\pi k^{+}} \mathbf{k}^{4}\left(\xi(1-\xi) \mathbf{k}^{2}+\tilde{\mathbf{p}}^{2}+m^{2}\right) \xi(1-\xi)d}\left|g_{i}^{a}(k)\right\rangle \nonumber\\
&&= -\int_{\bf x z} \int_{\Lambda}^{e^{\delta Y} \Lambda} d k^{+} \frac{g^{3} \rho^{a}(\mathbf{x})}{32 \pi^{{7/2}} \sqrt{k^{+}}}\Bigg\{ \frac{-i X_{i}}{\mathbf{X}^{2}} \left[\frac{2}{3}\left(-\frac{2}{\epsilon}+\ln \frac{m^{2}}{\bar{\mu}^{2}}\right)+g_{0}(m|\mathbf{X}|)\right.  \nonumber\\
&&\left. \quad+\frac{1}{3}\left(1-\frac{m^{2}}{m^{2}-\delta^{2}}\right)\right]+im^2X_iK_0[\delta|\mathbf{X}|]\left[\left(-\frac{2}{\epsilon}-1\right)+\ln \frac{m^{2}-\delta^{2}}{\bar{\mu}^2} \right] \Bigg\}|g_i^a(k^+,\mathbf{z})\rangle
\label{eq:psig1prime-final}
\end{eqnarray}
where we had to introduce $\delta$ as an IR regulator, modifying the $1/\mathbf{k}^4$ pole in the second line of Eq.~(\ref{eq:psig1prime-final}) to $1/(\mathbf{k}^2+\delta^2)^2$. $K_0$ is logarithmically divergent as $\delta \to 0$,  and
\begin{equation}
	g_{0}(x)\equiv\int_{0}^{1} d \xi\left(\left(4 \xi^{2}-4 \xi+2\right) K_{0}\left[\frac{x}{\sqrt{\xi(1-\xi)}}\right]\right).
\end{equation}
The IR divergence which we encounter is a furious divergence due to use of the LC quantization. 
This divergence gets canceled by introduction of a mass dependent counter term (for a more detailed discussion, consult Refs. \cite{Zhang:1993dd, Zhang:1993is, Harindranath:1993de}). The mass dependent counter term is introduced in order to ensure that the gluon remains massless. In other words, the renormalization condition is absence of any mass corrections to the gluon mass, to all orders in perturbation theory.

The dispersion relation on the light-cone coordinate is 
\begin{equation}
	m^2 = k^+k^- - \mathbf{k}_{\perp}^2.
\end{equation}
The gluon's pole mass $m^2$ has to vanish.
At fixed $k^+$ and $\mathbf{k}_\perp$,  the mass correction $\delta m^2$ is proportional to the LC  energy correction $\delta E_g(k)$, 
\begin{equation}
	\delta E_g(k) = \frac{\delta m^2}{2 k^+}.
\end{equation}
So, vanishing of $\delta m^2$ is equivalent to vanishing of $\delta E_g(k)$. 
\begin{figure}[htb]
	\centering
	\includegraphics[width=14cm]{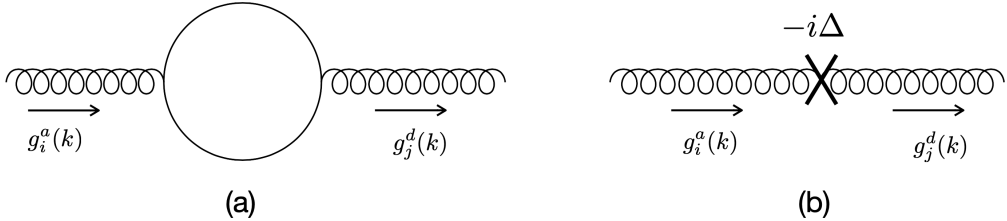}
	\caption{Self energy diagram (a) and the counter term (b) needed to ensure  vanishing of the 
	perturbative correction to the gluon mass.}
	\label{fig:self-energy}
\end{figure}
The  LC energy (self-energy) correction at one loop (order ${\cal O}(g^2)$)
 shown in Fig.~\ref{fig:self-energy}~(a) is
\begin{eqnarray}
\delta E_g(k)  & = & -\int d p^{+} d q^{+} \int d^2\mathbf{p} d^{2} \mathbf{q} \langle g_{j}^{d}({r})\left|H_{g q q}\right| q_{\lambda_{1}}^{\alpha}(p){\bar{q}}_{\lambda_{2}}^{\beta}{(q)}\rangle \langle q_{\lambda_{1}}^{\alpha}(p) \bar{q}_{\lambda_{2}}^{\beta}(q) | H_{gqq} | g_i^{a}(k)
 \rangle \nonumber\\
 && \times \frac{1}{E_{q\bar{q}}(p,q)-E_g(k)} \nonumber\\
 &=& -\frac{2 g^{2} \operatorname{tr}\left[t^{d} t^{a}\right]}{8 \pi k^{+}} \delta^{3}(r-k) \delta_{i j} \int_{0}^{1} d \xi \frac{1}{\xi(1-\xi) d} \int \frac{d^{d} \tilde{\mathbf{p}} }{(2\pi)^d} \frac{\left(4 \xi^{2}-4 \xi+d\right) \mathbf{p}^{2}+m^{2} d}{\tilde{\mathbf{p}}^{2}+m^{2}} \nonumber\\
 \label{eq:delta-p}
\end{eqnarray}
Derivation of Eq.~(\ref{eq:delta-p}) is nearly identical to that of Eq.~(\ref{eq:psi-g1}) (see Appendix~\ref{app:psig-1}) except for the difference in the energy denominators. The denominator in $\delta E_g(k)$
\begin{eqnarray}
	\frac{1}{E_{q\bar{q}}(p,q)-E_g(k)} &=& \frac{1}{\frac{\mathbf{p}^{2}+m^{2}}{2 p^{+}}+\frac{(\mathbf{k}-\mathbf{p})^{2}+m^{2}}{2\left(k^{+}-p^{+}\right)}-\frac{\mathbf{k}^{2}}{2 k^{+}}} =\frac{2 \xi(1-\xi) k^{+}}{\tilde{\mathbf{p}}^{2}+m^{2}}
\end{eqnarray}
where, again, $\xi \equiv p^+/k^+$ and $\tilde{\mathbf{p}} \equiv \mathbf{p}-\xi \mathbf{k}$.
In order to impose $\delta m^2 = 0$, 
the following mass dependent  counter term (Fig.~\ref{fig:self-energy}~(b))
 is added to the effective LC Hamiltonian
 \begin{equation}
	H_{gg}^{{c.t.}}=\frac{1}{2} \Delta (A_i^a)^2,
	\label{eq:ct-hamiltonian}
\end{equation}
where
\begin{equation}
	\Delta=\frac{g^{2}}{4 \pi} \int_{0}^{1} d \xi \frac{1}{\xi(1-\xi) d} \int \frac{d^d \tilde{\mathbf{p}} }{(2\pi)^d} \frac{\left(4 \xi^{2}-4 \xi+d\right) \tilde{\mathbf{p}}^{2}+m^{2} d}{\tilde{\mathbf{p}}^{2}+m^{2}}.
	\label{eq:counter-term-Delta}
\end{equation}
While it is not obvious from Eq.~(\ref{eq:counter-term-Delta}), $\Delta$ is proportional to $m^2$.
In dimensional regularization,
\begin{equation}
	\Delta = \frac{g^{2} m^{2}}{8 \pi^{2}}\left(\frac{2}{\epsilon}+\ln \frac{\bar{\mu}^{2}}{m^{2}}+1\right).
	\label{eq:Delta-final}
\end{equation}
The counter term ~(\ref{eq:ct-hamiltonian}) contributes to the one gluon component of the wave function 
\begin{eqnarray}
	\left|\psi_{g}^{1}\right\rangle_{c . t .} &\equiv& \int_{\Lambda}^{e^{\delta Y}\Lambda} d k^{+} d r^{+} \int d^{2} \mathbf{k} d^{2} \mathbf{r} \left|g_{j}^{d}({r})\right\rangle\left\langle g_{j}^{d}(r)\left|H_{g g}^{c.t.}\right| g_{i}^{a}(k)\right\rangle\left\langle g_{i}^{a}(k)\left|H_{g}\right| 0\right\rangle \nonumber \\
	&&\times\frac{1}{E_g(r)E_g(k)} 
	\label{eq:ct-wavefunction}
\end{eqnarray}
Using the matrix element
\begin{equation}
	\left\langle g_{j}^{d}(r)\left|H_{g g}^{c . t .}\right| g_{i}^{a}(k)\right\rangle = \Delta \frac{\delta^{i j} \delta^{a d}}{2 k^{+}} \delta^{3}(r-k),
\end{equation}
it is clear that the  energy denominator $1/(E_g(r)E_g(k))$ give rise to $1/\mathbf{k}^4$  pole, which cancels the similar pole  in Eq.~(\ref{eq:psig1prime-final}). Introducing
the IR regulator $\delta$, as was done in Eq.~(\ref{eq:psig1prime-final}), one immediately obtains (with the help of Eq.~(\ref{eq:int-gluon-mass-reg-0}))
\begin{eqnarray}
	\left|\psi_{g}^{1}\right\rangle_{c . t .} &=&\frac{g^{3} m^{2}}{32 \pi^{7 / 2}} \int_{\Lambda}^{e^{\delta Y} \Lambda} \frac{d k^{+}}{\sqrt{k^{+}}} \int_{\mathbf{x} \mathbf{z}} \rho^{a}(\mathbf{x}) \left(-i X_{i}\right) K_{0}\left[\delta |{X}|\right]\left(\frac{2}{\epsilon}-\ln \frac{m^{2}}{\bar{\mu}^{2}}+1\right)\left|g_{i}^{a}(z)\right\rangle \nonumber \\
	\label{eq:ct-wavefunction-final}
\end{eqnarray}
Combining the perturbative result with the counter term
\begin{equation}
\left|\psi_{g}^{1}\right\rangle=\left|\psi_{g}^{1^\prime}\right\rangle+\left|\psi_{g}^{1}\right\rangle_{c . t .},
\end{equation}
exact cancellation of the IR divergent terms (the terms proportional to $K_0[\delta |{X}|]$) is clearly observed.

In fact, introduction of the IR regulator is not necessary, if  Eq.~(\ref{eq:counter-term-Delta})  is used
for the counter term in its integral form,  instead of  the expression ~(\ref{eq:Delta-final}). Adding the counter term ~(\ref{eq:ct-wavefunction}) to the one gluon component ~(\ref{eq:psi-g1}) prior to momentum integration, 
we obtain
\begin{eqnarray}
|\psi_g^1\rangle&=&-\int_{\Lambda}^{e^{\delta Y} \Lambda} d k^{+} \int \frac{d^{2} \mathbf{k}}{(2 \pi)^{2}} \frac{d^{d} \tilde{\mathbf{p}}}{(2 \pi)^{d}} \int_{0}^{1} d \xi \frac{g^{3} \rho^{a}(-\mathbf{k}) k_{i}}{2 \sqrt{\pi k^{+}} \mathbf{k}^{4}} \frac{1}{\xi(1-\xi) d}\left[\left(4 \xi^{2}-4 \xi+d\right) \tilde{\mathbf{p}}^{2}+m^{2} d\right]\nonumber\\
&&\times \left(\frac{1}{\xi(1-\xi) \mathbf{k}^{2}+\tilde{\mathbf{p}}^{2}+m^{2}}-\frac{1}{\tilde{\mathbf{p}}^{2}+m^{2}}\right)|g_i^a(k)\rangle.
\label{eq:psi-g1-with-ct}
\end{eqnarray}
The second line of Eq.~(\ref{eq:psi-g1-with-ct}) is
\begin{equation}
	\left(\frac{1}{\xi(1-\xi) \mathbf{k}^{2}+\tilde{\mathbf{p}}^{2}+m^{2}}-\frac{1}{\tilde{\mathbf{p}}^{2}+m^{2}}\right)=-\frac{\xi(1-\xi) \mathbf{k}^{2}}{(\xi(1-\xi) \mathbf{k}^{2}+\tilde{\mathbf{p}}^{2}+m^{2})(\tilde{\mathbf{p}}^{2}+m^{2})}.
	\label{eq:addtion-k-squared}
\end{equation}
The  factor $\mathbf{k}^2$ in the numerator of Eq.~(\ref{eq:addtion-k-squared})  makes the $\mathbf{k}$-integral in Eq.~(\ref{eq:psi-g1-with-ct}) IR finite.
The $\tilde{\mathbf{p}}$-integral in Eq.~(\ref{eq:psi-g1-with-ct}) can be carried out using the typical Feynman parameter, which gives
\begin{eqnarray}
	&&\mu^{\epsilon} \int \frac{d^{d}{\tilde{\mathbf{p}}}}{(2 \pi)^{d}} \frac{1}{d} \frac{\left(4 \xi^{2}-4 \xi+d\right) \tilde{\mathbf{p}}^{2}+m^{2} d}{\left(\xi(1-\xi) \mathbf{k}^{2}+\tilde{\mathbf{p}}^{2}+m^{2}\right)\left(\tilde{\mathbf{p}}^{2}+m^{2}\right)} \nonumber \\
	&&=\frac{1}{8 {\pi}} \int_{0}^{1} d t\left[\left(\frac{2}{\epsilon}-\ln \frac{t \xi(1-\xi) \mathbf{k}^{2}+m^{2}}{\bar{\mu}^{2}}\right)\left(4 \xi^{2}-4 \xi+2\right)+\frac{2 m^{2}}{t \xi(1-\xi) \mathbf{k}^{2}+m^{2}}-2\right]. \nonumber \\
	{} \label{eq:loop-integral-p}	
\end{eqnarray}
And then the inverse Fourier Transform of $\mathbf{k}$-integral in Eq.~(\ref{eq:psi-g1-with-ct}) back to $\mathbf{x}, \mathbf{z}$ is straightforward using the integrals collected in Appendix \ref{app:integrals}. The result reads

\begin{eqnarray}
| \psi_g^1  \rangle	 = \int_{\Lambda}^{e^{\delta Y}\Lambda} \frac{d k^{+}}{\sqrt{k^{+}}} \int_{\mathbf{x} \mathbf{z}} \frac{g^{3} \rho^{a}(\mathbf{x})}{32 \pi^{3 / 2}} \frac{i {X}_{i}}{{X}^{2}} \left[\frac{2}{3}\left(-\frac{2}{\varepsilon}+\ln \frac{m^{2}}{\bar{\mu}^{2}}\right)+g_{1}(m|{X}|)\right]\left|g_{i}^{a}(k^+, \mathbf{z})\right\rangle,
\label{eq:psi-g1-final}
\end{eqnarray}
 where 
 \begin{equation}
 	g_1(x) \equiv \int_{0}^{1} d \xi \int_{0}^{1} d t\left\{2\left(2 \xi^{2}-2 \xi+1\right) K_{0}\left[\frac{x}{\sqrt{t \xi(1-\xi)}}\right]+\frac{x}{\sqrt{t \xi (1-\xi)}} K_{1}\left[\frac{x}{\sqrt{t\xi(1-\xi)}}\right]\right\}
 \end{equation}
The final result in~(\ref{eq:psi-g1-final}) is IR finite and has a well defined massless limit. 
Using small argument expansion of the Bessel functions,
\begin{equation}
	K_0[x]=-\gamma_E-\ln \frac{x}{2} + {\cal O}(x^2), \quad x K_1[x]=1+{\cal O}(x^2),
	\label{eq:bessel-taylor}
\end{equation}
we obtain
\begin{equation}
	g_1(x) = \frac{2}{3}(-2 \gamma_E -\ln \frac{x^2}{4})-\frac{10}{9}+{\cal O}(x^2),
	\label{eq:g1}
\end{equation}
The $\ln m^2$ in the expansion of $g_1(m|{X}|)$ is canceled  by the explicit $\ln(m^2/\bar{\mu}^2)$ term 
(\ref{eq:psi-g1-final}). This way,  the massless result ~\cite{Lublinsky:2016meo} is recovered.

\subsubsection{Subtraction of $K_{q\bar q}^m$}
  
In the massless case, the kernel $K_{q\bar{q}}^m$  (\ref{Kqq}) has a non-integrable  UV divergence when $\mathbf{z} \to \mathbf{z'}$, even though 
the full Hamiltonian is free of UV divergencies.  The same divergence appears in (\ref{eq:Kqq}),
since introducing masses for quarks does not modify the UV behavior. 

In the Hamiltonian, the UV divergence is regularized by subtracting the term (see the third line in Eq.~(\ref{eq:jimwlk-hamiltonian-massless}))
\begin{equation}
\int_{\mathbf{z}^{\prime}} K_{q \bar{q}}\left(\mathbf{x}, \mathbf{y}, \mathbf{z}, \mathbf{z}^{\prime}\right)J_{L}^{a}(\mathbf{x}) S_{A}^{a b}(\mathbf{z}) J_{R}^{b}(\mathbf{y})
\end{equation}

The very same term is then added back in the first line in Eq.~(\ref{eq:jimwlk-hamiltonian-massless}) (the real emission term)
which is then absorbed into the definition of $K_{JSJ}(\mathbf{x,y,z})$.  This contribution is given by
\begin{equation}
	\int_{\mathbf{z}^{\prime}} K_{q \bar{q}}\left(\mathbf{x}, \mathbf{y}, \mathbf{z}, \mathbf{z}^{\prime}\right).
\end{equation}
We postpone the discussion of $K_{JSJ}^m$ until the next subsection.  Meanwhile we focus on the subtraction.
Hence, our goal is to calculate the $\mathbf{z'}$-integral of Eq.~(\ref{eq:Kqq}). It turns out, however, that it is
 difficult to  carry out the integration directly. Instead, we start from Eq.~(\ref{eq:sigma-qq-after-tr}) and perform
  the $\mathbf{z'}$-integral first, before doing the Fourier transform (more details on the derivation are given in Appendix~\ref{app:subtractions}):
\begin{eqnarray}
	&& \int_{\mathbf{z'}} K^m_{q\bar{q}}(\mathbf{x,y,z,z'}) \nonumber \\
	&& 	=\frac{g^4}{(2\pi)^{10}} \int_{\mathbf{z}^{\prime}} \int_{0}^{1} d \xi \int d^2 \mathbf{k} d^{2} \mathbf{p} d^{2} \mathbf{u} d^{2}\mathbf{v} e^{-i \mathbf{v} \cdot {Z}+i \mathbf{u} \cdot {Y}^{\prime}+i \mathbf{p} \cdot {Z}-i \mathbf{k} \cdot {X}^{\prime}} \nonumber \\ 
	&& \quad \times \frac{1}{\left((1-\xi) \mathbf{v}^{2}+\xi(\mathbf{u}-\mathbf{v})^{2}+m^{2}\right)\left((1-\xi) \mathbf{p}^{2}+\xi(\mathbf{k}-\mathbf{p})^{2}+m^{2}\right)} \Bigg[\xi^{2}\nonumber \\
	&&\quad +\xi(1-2 \xi)\left(\frac{\mathbf{k} \cdot \mathbf{p}}{\mathbf{k}^{2}}+\frac{\mathbf{u} \cdot \mathbf{v}}{\mathbf{u}^{2}}\right) +\frac{(1-2 \xi)^{2} \mathbf{u} \cdot \mathbf{v} \cdot \mathbf{k} \cdot \mathbf{p}+\mathbf{u} \cdot \mathbf{k} \mathbf{v} \cdot \mathbf{p}-\mathbf{u} \cdot \mathbf{p} \mathbf{v} \cdot \mathbf{k}+m^{2} \mathbf{u} \cdot \mathbf{k}}{\mathbf{u}^{2} \mathbf{k}^{2}} \Bigg] \nonumber \\
	&& = \frac{g^4}{(2\pi)^4} \int_0^1 d\xi \int \frac{d^2\mathbf{k}}{2\pi} \frac{d^2\mathbf{p}}{(2\pi)^2} \frac{d^2\mathbf{u}}{2\pi} e^{-i \mathbf{k}\cdot X + i \mathbf{u}\cdot Y} \frac{1}{\mathbf{u}^2\mathbf{k}^2}  \nonumber \\     
	 &&\quad \times \Bigg[1+\frac{\mathbf{k} \cdot \mathbf{p}-\left(\mathbf{p}^{2}+m^{2}\right)}{(1-\xi)(\mathbf{k}-\mathbf{p})^{2}+\xi \mathbf{p}^{2}+m^{2}} +\frac{\mathbf{u} \cdot \mathbf{p}-\left(\mathbf{p}^{2}+m^{2}\right)}{(1-\xi)(\mathbf{u}-\mathbf{p})^{2}+\xi \mathbf{p}^{2}+m^{2}} \nonumber \\
	 &&\quad+\frac{\left(\mathbf{p}^{2}+m^{2}\right)(\mathbf{u}-\mathbf{p}) \cdot(\mathbf{k}-\mathbf{p})+\left(\mathbf{p}^{2}+m^{2}\right) m^{2}}{\left((1-\xi)(\mathbf{k}-\mathbf{p})^{2}+\xi \mathbf{p}^{2}+m^{2}\right)\left((1-\xi)(\mathbf{k}-\mathbf{p})^{2}+\xi \mathbf{p}^{2}+m^{2}\right)} \Bigg] 
	 \label{eq:subtraction-1}
\end{eqnarray}
The $\mathbf{z'}$-integral is now trivial and contributes a delta-function, which is then used to eliminate another integral,  the $\mathbf{v}$-integral.\footnote{In Eq.~(\ref{eq:subtraction-1}), the arrangement of the expression in this fashion is largely motivated by the massless expressions in Ref.~\cite{Kovchegov:2006wf}. \label{fn:algebra}} 
Since the phase factor of the Fourier transform does not involve $\mathbf{p}$, we first carry out the $\mathbf{p}$-integral:
\begin{eqnarray}
	&&\int_{\mathbf{z'}} K^m_{q\bar{q}}(\mathbf{x,y,z,z'}) \nonumber \\
	&&=\frac{g^{4}}{32 \pi^{5}} \int_{0}^{1} d \alpha \alpha(1-\alpha) \int \frac{d^{2} \mathbf{k}}{2 \pi} \frac{d^{2} \mathbf{u}}{2 \pi} e^{-i \mathbf{k} \cdot X+i \mathbf{u} \cdot Y} \Bigg[ -\frac{\mathbf{u} \cdot \mathbf{k}}{\mathbf{u}^{2} \mathbf{k}^{2}} \ln \frac{\alpha(1-\alpha)(\mathbf{u}-\mathbf{k})^{2}+m^{2}}{\bar{\mu}^{2}} \nonumber \\
	&& \quad + \frac{1}{\mathbf{u}^{2}} \ln \frac{\alpha(1-\alpha)(\mathbf{u}-\mathbf{k})^{2}+m^{2}}{\alpha(1-\alpha) \mathbf{k}^{2} + m^2}+\frac{1}{\mathbf{k}^{2}} \ln \frac{\alpha(1-\alpha)(\mathbf{u}-\mathbf{k})^{2}+m^{2}}{\alpha(1-\alpha) \mathbf{u}^{2}+m^2}
	 \Bigg] \nonumber \\
	 &&= \frac{g^4}{32\pi^5} \int_0^1 d \alpha \alpha(1-\alpha) \Bigg\{ \frac{2 X \cdot Y}{X^{2} Y^{2}}\left(-\ln[\alpha(1-\alpha)] - 2 K_{0}\left[\frac{m|X|}{\sqrt{\alpha(1-\alpha)}}\right]-\ln \frac{m^{2}}{\alpha(1-\alpha)\bar{\mu}^{2}}\right)\nonumber \\
	 && \quad+2F\left[\frac{m}{\sqrt{\alpha(1-\alpha)}},-X, Y\right] 
	 + \frac{1}{X^{2}} \ln \left[\frac{(X-Y)^{2}}{Y^{2}}\right] \frac{m|X|}{\sqrt{a(1-\alpha)}} K_{1}\left[\frac{m|X|}{\sqrt{\alpha(1-\alpha)}}\right] \nonumber \\
	 && \quad +\frac{1}{Y^{2}} \ln \left[\frac{(X-Y)^{2}}{X^{2}}\right] \frac{m|Y|}{\sqrt{\alpha(1-\alpha)}} K_{1}\left[\frac{m|Y|}{\sqrt{\alpha(1-\alpha)}}\right]\Bigg\},
	 \label{eq:subtraction} 	  
\end{eqnarray}
where 
$F[m,\mathbf{x},\mathbf{y}]$ is defined in Eq.~(\ref{eq:F-def}), and its massless limit is given in Eq.~(\ref{eq:F-massless-limit}). To derive the last equality of Eq.~(\ref{eq:subtraction}), Eqs.~(\ref{eq:double-Fourier-one-over-ksquared}) and (\ref{eq:double-integrals-kdotp-mu}) have been used.  It is then straightforward (using Eq.~(\ref{eq:bessel-taylor})) to show that as $m \to 0$,
\begin{eqnarray}
\int_{\mathbf{z'}} K^m_{q\bar{q}}(\mathbf{x,y,z,z'}) &\to& 
-\frac{g^{4}}{192 \pi^{5}}\left[\frac{X \cdot Y}{X^{2} Y^{2}}\left(\frac{10}{3}-4 \gamma_{E}+2\ln 4\right)+\frac{(X-Y)^{2}}{X^{2} Y^{2}} \ln \frac{(X-Y)^{2}}{X^{2} Y^{2} \bar{\mu}^{2}}\right. \nonumber \\
&&\left.+\frac{1}{X^{2}} \ln \left(\bar{\mu}^{2} X^{2}\right)+\frac{1}{Y^{2}} \ln \left(\bar{\mu}^{2} Y^{2}\right)\right],
\end{eqnarray}
which reproduces Eq.~(D.5) of Ref.~\cite{Lublinsky:2016meo}.

\subsubsection{Extracting $K_{JSJ}^m$}
With the NLO wave function Eq.~(\ref{eq:psi-g1-final}) and the subtraction term Eq.~(\ref{eq:subtraction}) at hand, it is straightforward to extract the  evolution kernel $K_{JSJ}^{(m)}$. First, we compute the matrix element
\begin{eqnarray}
	&&\Sigma_{JSJ}^{\text{NLO}} \equiv \left\langle\psi_{g}^{1}\left|\hat{S} \right| \psi_{g}^{\text{LO}}\right\rangle + \left\langle\psi_{g}^{\text{LO}}\left|\hat{S} \right| \psi_{g}^{1}\right\rangle \nonumber \\
	&& = \int_{\Lambda}^{e^{\delta{Y}} \Lambda} \frac{d k^{+}}{\sqrt{k^{+}}} \int_{\mathbf{x} \mathbf{z}} \frac{g^{3} \rho^{a}(\mathbf{x})}{32 \pi^{7 / 2}} \frac{-i X_{i}}{X^{2}}\left[\frac{2}{3}\left(-\frac{2}{\epsilon}+\ln \frac{m^{2}}{\bar{\mu}^{2}}\right)+g_{1}(m|X|)\right]\left\langle g_{i}^{a}\left(k^{+}, \mathbf{z}\right)\right| \nonumber \\
	&&\quad \times (\hat{S}-1) \int_{\Lambda}^{e^{\delta Y}\Lambda} \frac{d k^{\prime{+}}}{\sqrt{k^{\prime+}}} \int_{\mathbf{y}, \mathbf{z}^{\prime}} \frac{i g Y^{\prime i^{\prime}}}{2 \pi^{3 / 2} Y^{\prime 2}} \rho^{a^{\prime}}(\mathbf{y})\left|g_{i^{\prime}}^{a^{\prime}}\left(k^{\prime{+}}, \mathbf{z}^{\prime}\right)\right\rangle + (\psi_g^{1} \leftrightarrow \psi_g^{\text{LO}}) \nonumber \\
	&&= \delta Y \frac{g^{4}}{64 \pi^{5}} \int_{\mathbf{x} \mathbf{y} \mathbf{z}} \frac{X \cdot Y}{X^{2} Y^{2}} J_{L}^{a}(\mathbf{x}) S^{a b} J_{R}^{b}(\mathbf{y}) \left[\frac{2}{3}\left(-\frac{2}{\epsilon}+\ln \frac{m^{2}}{\bar{\mu}^{2}}\right)+g_{1}(m | X|)\right] + (\psi_g^{1} \leftrightarrow \psi_g^{\text{LO}})\nonumber \\	
	&&= \delta Y \frac{g^{4}}{64 \pi^{5}} \int_{\mathbf{x} \mathbf{y} \mathbf{z}} \frac{X \cdot Y}{X^{2} Y^{2}} J_{L}^{a}(\mathbf{x}) S^{a b} J_{R}^{b}(\mathbf{y}) \left[\frac{4}{3}\left(-\frac{2}{\epsilon}+\ln \frac{m^{2}}{\bar{\mu}^{2}}\right)+g_{1}(m | X|) +g_{1}(m | Y|)\right], \nonumber \\
\end{eqnarray}
where $| \psi_g^{\rm LO}\rangle$ is the LO one-gluon emission wave function given in Eq.~(\ref{eq:psig-LO}), and $g_1(x)$ is defined in Eq.~(\ref{eq:g1}). Then, discarding the $1/\epsilon$ term, the evolution kernel reads 
\begin{eqnarray}
K_{JSJ}^{\prime m}=\frac{g^{4}}{128 \pi^{5}}  \frac{X \cdot Y}{X^{2} Y^{2}}  \left[\frac{4}{3}\ln \frac{m^{2}}{\bar{\mu}^{2}}+g_{1}(m | X|) +g_{1}(m | Y|)\right].
\label{eq:KJSJ-prime}
\end{eqnarray}
where $1/128$ comes from $1/(64\times2)$ due to the convention that $K_{JSJ}$ corresponds to the coefficient of $2 J_{L}^{a} S^{a b} J_{R}^{b} $. In the massless limit of $g_1(m|X|)$ and $g_1(m|Y|)$, Eq.~(\ref{eq:KJSJ-prime}) reproduces the corresponding  massless limit  of Ref.~\cite{Lublinsky:2016meo}  (Eq.~(4.36) therein). As discussed previously, the final result 
 $K^m_{JSJ}$ is obtained after subtraction 
\begin{equation}
	K_{JSJ}^{m} = K_{JSJ}^{\prime m}-\frac{1}{2}\int_{\mathbf{z'}}K^m_{q\bar{q}}(\mathbf{x,y,z,z'})
	\label{eq:KJSJ-final}
\end{equation}
with the subtraction contribution given in Eq.~(\ref{eq:subtraction}). 
We note in passing that part of our result  can be identified with the mass effects in the running coupling, since the coupling running is related to the resummation of the quark loop diagram in Fig.~\ref{fig:self-energy}. We, however, don't see it instructive to separate these effects from the rest of the kernel.

Finally, we would like to remark on the virtual terms of the type $J_{L}^{a}  J_{L}^{a} $. 
They are expected to come with the same kernel $K_{JSJ}^{m} $, though we have  not explicitly demonstrated
this  here. These terms appear from the normalization of the wave function. 
 It is straightforward to see \cite{Lublinsky:2016meo} that the kernel in front of the virtual terms 
 is indeed the same as the one in front of the real term, including the mass corrections.

\subsubsection{CWZ subtraction scheme }
We would like to comment more on the subtraction scheme used in our work.\footnote{The authors thank the referee for the suggestion that we comment on the use of CWZ subtraction scheme in our current study.} So far, we have been using $\overline{\rm MS}$ subtraction scheme for the massive quarks, which means a quadratic gluon counter terms are introduced to cancel the $1/\epsilon$ UV divergence in Eq.~(\ref{eq:psi-g1-final}). Note that the $1/\epsilon$ canceling counter term (which exists also for diagrams with massless quarks) is in addition to the mass dependent counter term in Figure~\ref{fig:self-energy} (b). When quark masses are relevant in a physical process, it is also useful to use the Collins-Wilczek-Zee (CWZ) subtraction scheme \cite{Collins:1978wz,Qian:1984kf,Collins:1986mp,Collins:1998rz}, which makes the decoupling of heavy quark manifest in the large mass limit. In this subtraction scheme, the ordinary $\overline{\rm MS}$ subtraction is carried out for diagrams with active quarks (light quarks) and gluons, while for inactive quarks (heavy quarks) zero momentum subtractions are implemented.



\begin{figure}[htb]
	\centering
	\includegraphics[width=15cm]{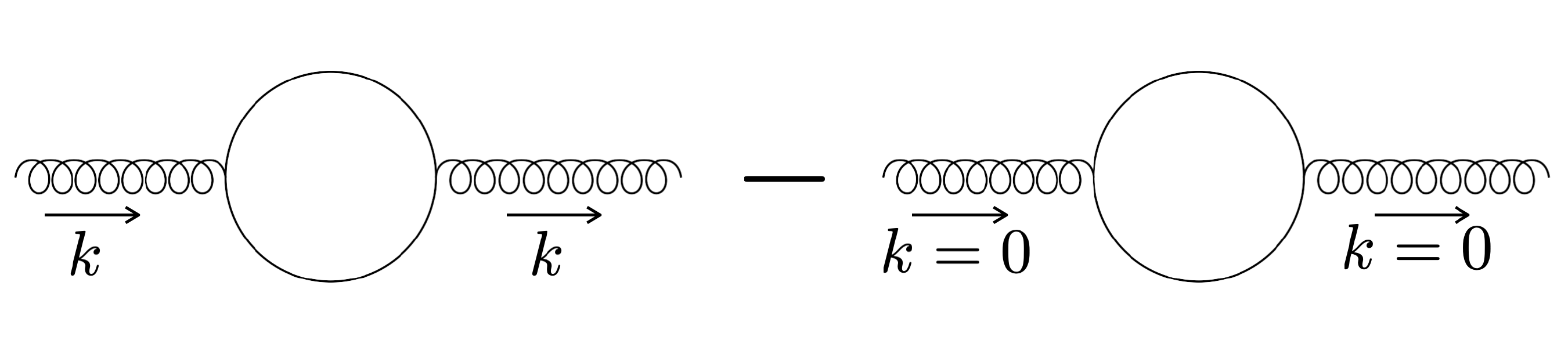}
	\caption{Zero momentum subtraction of the sub-diagram (meaning the gluon self-energy diagram) implemented to calculated the wave function due to loop diagram in Figure~\ref{fig:quark-loop-diagram}.}
	\label{fig:cwz}
\end{figure}

Figure~\ref{fig:cwz} shows how zero momentum subtraction is carried out for the gluon vacuum polarization: the loop integral with external momentum $k \to 0$ is subtracted from the original loop integral. This vacuum polarization diagram is a sub-diagram for the loop correction diagram as shown in Figure~\ref{fig:quark-loop-diagram}. So a zero momentum subtraction should be implemented for Eq.~(\ref{eq:loop-integral-p}). By setting $k \to 0$, one obtains for the wave function
\begin{eqnarray}
	&&\left|\psi_{g}^{1}\right\rangle_{0-\rm mom}= \nonumber\\
	&&-\int_{\Lambda}^{e^{\delta Y} \Lambda} d k^{+} \int \frac{d^{2} \mathbf{k}}{(2 \pi)^{2}} \int_{0}^{1} d \xi \frac{g^{3} \rho^{a}(-\mathbf{k}) k_{i}}{2 \sqrt{\pi k^{+}} \mathbf{k}^{2}} \frac{1}{8\pi}\left(\frac{2}{\epsilon}-\ln \frac{m^{2}}{\bar{\mu}^{2}}\right)\left(4 \xi^{2}-4 \xi+2\right) |g_i^a(k)\rangle \nonumber \\
	{} \label{eq:psi-g-zero-mom}
\end{eqnarray}
which is to be subtracted from Eq.~(\ref{eq:psi-g1-with-ct}). The zero momentum subtracted wave function is then simply obtained by replacing $\bar{\mu}$ by $m$ and removing $\frac{2}{\epsilon}$, as is usually the case for CWZ subtraction in ordinary equal time quantization. Note that since the loop integral (i.e., Eq.~(\ref{eq:loop-integral-p}) and Figure~\ref{fig:cwz}) is independent of $k^+$, i.e., the $k^+$ component does not flow into the loop, the zero momentum subtraction simply means zero $k_\perp$ momentum subtraction.


Similar argument shows that the CWZ implementation for $\int d \mathbf{z}^\prime K^{m}_{q\bar{q}}(\mathbf{x,y,z,z'})$ is equivalent to replacing $\bar{\mu} \to m$.  After replacing $\bar{\mu} \to m$ in Eq.~(\ref{eq:subtraction}) and Eq.~(\ref{eq:KJSJ-prime}) one obtains the kernel $K_{JSJ}^{m}$ in the CWZ subtraction scheme. Note that this removes all the $\log(m)$ dependence in $K_{JSJ}^{m}$. The final expressions for $K_{JSJ}^{m}$ in Eq.~(\ref{eq:KJSJ-final}) and $K_{q\bar{q}}^m$ in Eq.~(\ref{eq:Kqq}) have a modified Bessel function in every term (either $K_0[m|X|]$ or $K_1[m|Y|]$ for instance), which in the heavy mass limit are exponentially suppressed, so that the decoupling of heavy quark is manifest in the $m\to \infty$ limit. 

In the above, we have done the zero momentum subtraction for massive quarks in all external momentum region (i.e., $|\mathbf{k}| \in (0, \infty)$ in Eq.~(\ref{eq:loop-integral-p})), which shows the explicit decoupling of massive quarks in the large mass limit, and which is also a good cross check of our computations. In practice, however, we should carry out the CWZ subtraction of Eq.~(\ref{eq:loop-integral-p}) step by step when multiple quarks with different masses are involved (similar to the running coupling that should be matched step by step in the CWZ scheme). So with multiple massive quarks involved, after CWZ subtraction, Eq.~(\ref{eq:loop-integral-p}) should be replaced with a sum of step functions as follows
\begin{eqnarray}
	\text{Eq.~(\ref{eq:loop-integral-p})} \to &&\sum_i \frac{1}{8 {\pi}}\theta(m_i^2-\mathbf{k}^2)  \int_{0}^{1} d t\left[\left(-\ln \frac{t \xi(1-\xi) \mathbf{k}^{2}+m_i^{2}}{m_i^{2}}\right)\left(4 \xi^{2}-4 \xi+2\right) \right. \nonumber \\
	&&\left.+\frac{2 m_i^{2}}{t \xi(1-\xi) \mathbf{k}^{2}+m_i^{2}}-2\right] \nonumber \\
	&& + \frac{1}{8 {\pi}}\theta(\mathbf{k}^2-m_i^2)  \int_{0}^{1} d t\left[\left(-\ln \frac{t \xi(1-\xi) \mathbf{k}^{2}+m_i^{2}}{\bar{\mu}^{2}}\right)\left(4 \xi^{2}-4 \xi+2\right) \right. \nonumber \\
	&&\left.+\frac{2 m_i^{2}}{t \xi(1-\xi) \mathbf{k}^{2}+m_i^{2}}-2\right],
	{} \label{eq:cwz-steps}	
\end{eqnarray}
which is then plugged back in Eq.~(\ref{eq:psi-g1-with-ct}). Similar manipulations should be done for the subtraction contribution of $K^m_{JSJ}$, i.e., $\int_\mathbf{z'} K^m_{q\bar{q}}(\mathbf{x,y,z,z'})$. However, the Fourier Transforms back to coordinate space (the counter parts of Eq.~(\ref{eq:psi-g1-final}) and Eq.~(\ref{eq:subtraction})) become complex, and we do not discuss them further in the current work. 

\section{Conclusions}
\label{sec:conclusions}
In this paper, we generalized the work of Ref.~\cite{Lublinsky:2016meo},  exploring
 high energy evolution with non-vanishing quark masses.  We computed the  hadronic light-cone wave function  of a fast moving projectile, focusing on the effects originating from non-vanishing quark  masses. As an application of the wave function, we derived the NLO evolution kernels for the 
JIMWLK Hamiltonian, particularly the kernels $K^m_{q\bar{q}}$ and $K^m_{JSJ}$.  Both enter
the dipole evolution at NLO \cite{Balitsky:2007feb} through the form presented in \cite{Kovner:2013ona}.

There are only two types of diagrams that involve  massive quarks other than those that have already been calculated  in Ref.~\cite{Lublinsky:2016meo}: the quark splitting diagram (Fig.~\ref{fig:qq-diagram}) and the virtual quark loop diagram for one gluon component of the wave function (Fig.~\ref{fig:quark-loop-diagram}). At every step, we  carefully checked that the massless limit of Ref.~\cite{Lublinsky:2016meo} is recovered.

While a priori inclusion of  mass  looks like a very straightforward generalization of the previous calculations, we have encountered several complications.  First of all, on the technical level, some integrals became too complex to  be performed analytically. Furthermore, 
for the quark loop diagram, a new IR divergence (divergence coming from small $\mathbf{k}_\perp$) emerges. This is somewhat surprising since  masses generally cure IR divergences and not the other way around. In fact, this IR divergence is, to some extent, a known "artifact" of the LC quantization. 
It turns out that the cancellation of this divergence is realized by introducing a mass dependent gauge field counter term, which is required for the gluon to remain 
massless to all orders in perturbation  theory \cite{Zhang:1993dd, Zhang:1993is, Harindranath:1993de}. 

Unsurprisingly, we have not been able to obtain  simple fully analytical expressions for the kernels 
$K^m_{q\bar{q}}$ and $K^m_{JSJ}$ analogous to the ones of the massless limit. 
Our results for the kernels 
are expressed as one dimensional integrals. We are hopeful that they are nevertheless sufficiently simple and could be implemented
in future numerical studies  and  phenomenology.  Obviously, a complete 
NLO  calculation is already very challenging, particularly due to multiple coordinate integrations. 
In principle, one might expect that considering various quark mass limits (light/heavy limits) might lead to more tractable analytical expressions. We have considered both limits but have not  observed any simplifications worth reporting.  This is largely due to the absence of any clear scale separation between 
the mass  and momenta  which are integrated over in the kernels.


\acknowledgments

LD and ML were supported by the Binational Science Foundation grant \#2018722, 
by the Horizon 2020 RISE
 "Heavy ion collisions: collectivity and precision in saturation physics"  under grant agreement No. 824093, and 
 Horizon 2020 research and innovation program  under grant agreement STRONG – 2020 - No 824093.  
 LD was  partly supported by the Foreign Postdoctoral Fellowship Program of the Israel 
 Academy of Sciences and Humanities and  by the Alexander von Humboldt Foundation.

\appendix

\section{Computation of $\Sigma_{\bar{q} q}$ at NLO}
\label{app:psi-qq}
In this Appendix,  details on derivation of  Eq.~(\ref{eq:sigmaqq}) are provided. 
We start from Eq.~(\ref{eq:psi-qq-sum}) and the definition of  $\Sigma_{\bar{q} q}^{\rm NLO}$ in Eq.~(\ref{eq:sigma-qq-def}):
\begin{eqnarray}
&&\Sigma_{\bar{q} q}^{\rm NLO} = \nonumber\\
&&\left\langle \bar{q}_{\lambda_{4}}^{\delta}(u - v) q_{\lambda_{3}}^{\gamma}(v)\right|\Bigg\{ \sum_{\lambda_{3} \lambda_{4}} \int_{\Lambda}^{e^{\delta Y}\Lambda} d u^{+} \int_{0}^{1} d \theta \int \frac{d^{2} \mathbf{u}}{(2 \pi)^{2}} \frac{d^{2} \mathbf{v}}{(2 \pi)^{2}}  \frac{2 \pi g^{2} t_{\gamma \delta}^{b{*}} \rho^{b}(\mathbf{u}) \theta(1-\theta)}{(1-\theta) \mathbf{v}^{2}+\theta(\mathbf{u}-\mathbf{v})^{2}+m^{2}}
 \nonumber \\
 && \times  \chi_{\lambda_{4}}^{\dagger}\left[\frac{u_{i}}{\mathbf{u}^{2}}\left(\alpha_{i} \frac{\mathbf{\alpha} \cdot \mathbf{v}+\beta m}{\theta}+\frac{\mathbf{\alpha} \cdot(\mathbf{u}-\mathbf{v})-\beta{m}}{1-\theta} \alpha_{i}\right)\right] \chi_{\lambda_{3}}\Bigg\}   \nonumber\\
&& \times \hat{S} \Bigg\{ \sum_{\lambda_{1} \lambda_{2}} \int_{\Lambda}^{e^{\delta Y} \Lambda} d k^{+} \int_{0}^{1} d \xi \int \frac{d^{2} \mathbf{k}}{(2 \pi)^{2}} \frac{d^{2}\mathbf{p}}{(2 \pi)^{2}}\frac{2 \pi g^{2} t_{\alpha \beta}^{a} \rho^{a} (-\mathbf{k}) \xi(1-\xi)}{(1-\xi) \mathbf{p}^{2}+\xi(\mathbf{k}-\mathbf{p})^{2}+m^{2}} \nonumber \\
&& \times \chi_{\lambda_{1}}^{\dagger}\left[\frac{k_{i}}{\mathbf{k}^{2}}\left(\frac{\mathbf{\alpha} \cdot \mathbf{p}+\beta m}{\xi} \alpha_{i}+\alpha_{i} \frac{{\alpha} \cdot(\mathbf{k}-\mathbf{p})-\beta{m}}{1-\xi}\right)\right] \chi_{\lambda_{2}} \Bigg\} \left|\bar{q}_{\lambda_{2}}^{\beta}(k-p) q_{\lambda_{1}}^{\alpha}(p)\right\rangle \nonumber\\
\end{eqnarray}
where $\theta \equiv v^+/u^+$. After Fourier-transforming $\rho^b(\mathbf{u})$ and $\rho^a(-\mathbf{k})$ back to coordinate space (using Eq.~(\ref{eq:rho-fourier-conv})), plugging in Eq.~(\ref{eq:qqSqq}), and also converting $\rho^b(\mathbf{y}) \hat{S} \rho^a(\mathbf{x})$ $\to$ $J_R^b(\mathbf{y})  J_L^a(\mathbf{x})$ (see Sec.~\ref{sec:basics}), we get
\begin{eqnarray}
	&&\Sigma_{\bar{q} q}^{\rm NLO} \nonumber \\
	&&=\int_{\mathbf{x,y}} \Bigg\{ \sum_{\lambda_{3} \lambda_{4}} \int_{\Lambda}^{e^{\delta Y}\Lambda} d u^{+} \int_{0}^{1} d \theta \int \frac{d^{2} \mathbf{u}}{(2 \pi)^{2}} \frac{d^{2} \mathbf{v}}{(2 \pi)^{2}}  \frac{2 \pi g^{2} t_{\gamma \delta}^{b{*}} J_R^{b}(\mathbf{y}) \theta(1-\theta)}{(1-\theta) \mathbf{v}^{2}+\theta(\mathbf{u}-\mathbf{v})^{2}+m^{2}}
	\nonumber \\
	&& \quad\times  \chi_{\lambda_{4}}^{\dagger}\left[\frac{u_{i}}{\mathbf{u}^{2}}\left(\alpha_{i} \frac{\mathbf{\alpha} \cdot \mathbf{v}+\beta m}{\theta}+\frac{\mathbf{\alpha} \cdot(\mathbf{u}-\mathbf{v})-\beta{m}}{1-\theta} \alpha_{i}\right)\right] \chi_{\lambda_{3}}\Bigg\}   \nonumber\\
	&& \quad\times \Bigg\{ \sum_{\lambda_{1} \lambda_{2}} \int_{\Lambda}^{e^{\delta Y} \Lambda} d k^{+} \int_{0}^{1} d \xi \int \frac{d^{2} \mathbf{k}}{(2 \pi)^{2}} \frac{d^{2}\mathbf{p}}{(2 \pi)^{2}}\frac{2 \pi g^{2} t_{\alpha \beta}^{a} J_L^{a} (\mathbf{x}) \xi(1-\xi)}{(1-\xi) \mathbf{p}^{2}+\xi(\mathbf{k}-\mathbf{p})^{2}+m^{2}} \nonumber \\
	&& \quad\times \chi_{\lambda_{1}}^{\dagger}\left[\frac{k_{i}}{\mathbf{k}^{2}}\left(\frac{\mathbf{\alpha} \cdot \mathbf{p}+\beta m}{\xi} \alpha_{i}+\alpha_{i} \frac{{\alpha} \cdot(\mathbf{k}-\mathbf{p})-\beta{m}}{1-\xi}\right)\right] \chi_{\lambda_{2}} \Bigg\} \nonumber \\
	&& \quad\times \frac{\delta_{\lambda_{1} \lambda_{3}} \delta_{\lambda_{2} \lambda_{4}} }{(2 \pi)^{4} k^{+}} \int_{\mathbf{z}, \mathbf{z}^{\prime}}e^{-i \mathbf{v} \cdot (\mathbf{z}-\mathbf{z}^{\prime})+i \mathbf{p} \cdot\left(\mathbf{z}-\mathbf{z}^{\prime}\right)-i \mathbf{u} \cdot \mathbf{z}^{\prime}+i \mathbf{k} \cdot \mathbf{z}^{\prime}}  S^{\gamma \alpha}(\mathbf{z}) S^{\dagger\beta \delta}\left(\mathbf{z}^{\prime}\right) \delta\left(u^{+}-k^{+}\right) \delta(\xi-\theta)\nonumber\\	
	&&= \int_{\mathbf{x}, \mathbf{y}, \mathbf{z}, \mathbf{z}^{\prime}}  \int_{\Lambda}^{e^{\delta Y}\Lambda} d k^{+} \int_{0}^{1} d \xi \int_{\mathbf{k}, \mathbf{p}, \mathbf{u}, \mathbf{v}} e^{-i \mathbf{v} \cdot {Z}+i \mathbf{u} \cdot {Y}^{\prime}+i \mathbf{p} \cdot {Z}-i \mathbf{k} \cdot \mathbf{X}^{\prime}} \nonumber \\
	&&\quad\times \frac{g^{4} J_{L}^{a}(\mathbf{x}) {\rm tr}\left[S(z) t^{a} S^{\dagger}{\left(z^{\prime}\right)} t^{b}\right] J_{R}^{b}(\mathbf{y}) \xi^{2}(1-\xi)^{2}}{(2 \pi)^{10}\left((1-\xi) \mathbf{v}^{2}+\xi(\mathbf{u}-\mathbf{v})^{2}+m^{2}\right)\left((1-\xi) \mathbf{p}^{2}+\xi(\mathbf{k}-\mathbf{p})^{2}+m^{2}\right) k^{+}} \frac{1}{\mathbf{u}^{2} \mathbf{k}^{2}} \nonumber \\
	&& \quad\times 
		\operatorname{tr} \Bigg[ \left.\Lambda_{+}\left(\mathbf{\alpha} \cdot \mathbf{u} \frac{\mathbf{\alpha} \cdot \mathbf{v}+\beta m}{\xi}+\frac{\mathbf{\alpha} \cdot(\mathbf{u}-\mathbf{v})-\beta m}{1-\xi} \mathbf{\alpha} \cdot \mathbf{u}\right) \Lambda_{+}\right. \nonumber \\
&& \qquad \quad \quad \quad~		\left.\left(\frac{\mathbf{\alpha} \cdot \mathbf{p}+\beta m}{\xi} \mathbf{\alpha} \cdot \mathbf{k}+\mathbf{\alpha} \cdot \mathbf{k} \frac{\mathbf{\alpha} \cdot(\mathbf{k}-\mathbf{p})-\beta{m}}{1-\xi}\right)\right. \Bigg]
\label{eq:sigma-qq-tr}
\end{eqnarray}
where $Z, X', Y'$ are defined in Eq.~(\ref{eq:XYZ-convection}),
\begin{equation}
\int_{\mathbf{k,p,u,v}}\equiv \int d^2 \mathbf{k} d^2 \mathbf{p}d^2 \mathbf{u}d^2 \mathbf{v}, 
\end{equation}
 and $\sum_{\lambda} \chi_{\lambda} \chi_{\lambda}^{\dagger} = \Lambda_+$ is used to obtain the trace in the last equality.
After some Dirac algebra, the trace part becomes
\begin{eqnarray}
	&&\operatorname{tr} \Bigg[ \left.\Lambda_{+}\left(\mathbf{\alpha} \cdot \mathbf{u} \frac{\mathbf{\alpha} \cdot \mathbf{v}+\beta m}{\xi}+\frac{\mathbf{\alpha} \cdot(\mathbf{u}-\mathbf{v})-\beta m}{1-\xi} \mathbf{\alpha} \cdot \mathbf{u}\right)\right.\nonumber \\ 
	&& \quad ~\, \left. \Lambda_{+}\right. \left.\left(\frac{\mathbf{\alpha} \cdot \mathbf{p}+\beta m}{\xi} \mathbf{\alpha} \cdot \mathbf{k}+\mathbf{\alpha} \cdot \mathbf{k} \frac{\mathbf{\alpha} \cdot(\mathbf{k}-\mathbf{p})-\beta{m}}{1-\xi}\right)\right. \Bigg] \nonumber \\
&&	= \frac{2}{\xi^2(1-\xi)^2}\Big[\xi^{2} \mathbf{u}^{2} \mathbf{k}^{2}+\xi(1-2 \xi) \mathbf{u}^{2} \mathbf{k} \cdot \mathbf{p}+\xi(1-2 \xi) \mathbf{u} \cdot \mathbf{v} \mathbf{k}^{2}+(1-2 \xi)^{2} \mathbf{u} \cdot \mathbf{v} \mathbf{k} \cdot \mathbf{p}  \nonumber \\
&& \quad +\mathbf{u} \cdot \mathbf{k} \mathbf{v} \cdot \mathbf{p}-\mathbf{u} \cdot \mathbf{p} \mathbf{v} \cdot \mathbf{k}+m^{2} \mathbf{u} \cdot \mathbf{k} \Big] 
\label{eq:qq-tr}
\end{eqnarray}
Substituting Eq.~(\ref{eq:qq-tr}) back in Eq.~(\ref{eq:sigma-qq-tr}), we have
\begin{eqnarray}
	&&\Sigma_{\bar{q} q}^{\rm NLO}
	= \int_{\mathbf{x}, \mathbf{y}, \mathbf{z}, \mathbf{z}^{\prime}} \int_{\Lambda}^{e^{\delta Y}\Lambda} d k^{+} \int_{0}^{1} d \xi \int_{\mathbf{k}, \mathbf{p}, \mathbf{u}, \mathbf{v}} e^{-i \mathbf{v} \cdot {Z}+i \mathbf{u} \cdot {Y}^{\prime}+i \mathbf{p} \cdot {Z}-i \mathbf{k} \cdot {X}^{\prime}} \nonumber \\
	&&\times \frac{g^{4} J_{L}^{a}(\mathbf{x}) {\rm tr}\left[S(z) t^{a} S^{\dagger}{\left(z^{\prime}\right)} t^{b}\right] J_{R}^{b}(\mathbf{y})}{(2 \pi)^{10}\left((1-\xi) \mathbf{v}^{2}+\xi(\mathbf{u}-\mathbf{v})^{2}+m^{2}\right)\left((1-\xi) \mathbf{p}^{2}+\xi(\mathbf{k}-\mathbf{p})^{2}+m^{2}\right) k^{+}}\Bigg[\xi^{2}  \nonumber \\
	&& +\xi(1-2 \xi)\left(\frac{\mathbf{k} \cdot \mathbf{p}}{\mathbf{k}^{2}}+\frac{\mathbf{u} \cdot \mathbf{v}}{\mathbf{u}^{2}}\right)+\frac{(1-2 \xi)^{2} \mathbf{u} \cdot \mathbf{v} \mathbf{k} \cdot \mathbf{p}+\mathbf{u} \cdot \mathbf{k} \mathbf{v} \cdot \mathbf{p}-\mathbf{u} \cdot \mathbf{p} \mathbf{v} \cdot \mathbf{k}+m^{2} \mathbf{u} \cdot \mathbf{k}}{\mathbf{u}^{2} \mathbf{k}^{2}}\Bigg] \nonumber \\
	\label{eq:sigma-qq-after-tr}
\end{eqnarray}
which for massless quarks reproduces Eq.~(4.12) in Ref.~\cite{Lublinsky:2016meo}. Eq.~(\ref{eq:sigma-qq-tilde}) is obtained after 
 the following change of variables is implemented,
\begin{equation}
\mathbf{v} = \tilde{\mathbf{v}}+\xi \mathbf{u}, \quad \mathbf{p} = \tilde{\mathbf{p}}+\xi \mathbf{k}.
\end{equation}

\section{Computation of $| \psi_{g}^1 \rangle$ with quark loop correction}
\label{app:psig-1}
In this Appendix, we flash the missing steps in deriving Eq.~(\ref{eq:psig-1-final}) and (\ref{eq:ct-wavefunction-final}).
From Eq.~(\ref{eq:psi-g1}),
\begin{eqnarray}
	\label{eq:psig-1-Calc-1}
	\left|\psi_{g}^{1^\prime}\right\rangle &\equiv&-\int_{\Lambda}^{e^{\delta Y} \Lambda} d k^{+} d p^{+} d q^{+} d r^{+} \int d^{2} \mathbf{k} d^{2} \mathbf{p} d^{2} \mathbf{q} d^{2} \mathbf{r} \nonumber\\
	&& \times\left|g_{j}^{d}(r)\right\rangle \frac{\left\langle g_{j}^{d}(r)\left|H_{g q q}\right| q_{\lambda_{1}}^{\alpha}(p) \bar{q}_{\lambda_{2}}^{\beta}(q)\right\rangle\left\langle q_{\lambda_{1}}^{\alpha}(p) \bar{q}_{\lambda_{2}}^{\beta}(q)\left|H_{g q q}\right| g_{i}^{a}(k)\right\rangle\left\langle g_{i}^{a}(k)\left|H_{g}\right| 0\right\rangle}{E_{g}(r) E_{q \bar{q}}(p, q) E_{g}(k)}\nonumber\\
	\nonumber\\
	&=&\int_{\Lambda}^{e^{\delta Y}\Lambda}dk^{+}d p^{+} \int d^{2} \mathbf{k} d^{2} \mathbf{p} \frac{g^{3} \rho^{a}(-\mathbf{k}) \operatorname{tr}\left[t^{a} t^{d}\right] k_{i}}{32 \pi^{9 / 2} \sqrt{k^{+}} \mathbf{k}^{4}\left(\frac{\mathbf{p}^{2}+m^{2}}{p^{+}}+\frac{(\mathbf{k}-\mathbf{p})^{2}+m^{2}}{k^{+}-p^{+}}\right)} \nonumber\\
	&&\times~ \chi_{\lambda_{2}}^{\dagger}\left(\frac{2 k_{j}}{k^{+}}-\alpha_{j} \frac{\mathbf{\alpha} \cdot \mathbf{p}+\beta m}{p^{+}}-\frac{\mathbf{\alpha} \cdot(\mathbf{k}-\mathbf{p})-\beta m}{k^{+}-p^{+}} \alpha_{j}\right) \chi_{\lambda_{1}} \nonumber\\
	&&\times~ \chi_{\lambda_{1}}^{\dagger}\left(\frac{2 k_{i}}{k^{+}}-\frac{\mathbf{\alpha} \cdot \mathbf{p}+\beta m}{p^{+}} \alpha_{i}-\alpha_{i} \frac{\mathbf{\alpha} \cdot(\mathbf{k}-\mathbf{p})-\beta {m}}{k^{+}-p^{+}}\right) \chi_{\lambda_{2}} \left|g_{j}^{d}(k)\right\rangle
\end{eqnarray}
where the matrix elements ~(\ref{eq:Hgqq-ME}) and ~(\ref{eq:Hg-ME}) are used. The polarization indices $\lambda_{1}$ and $\lambda_{2}$ are summed over. The following part of the integrand
\begin{eqnarray}
	 &&\chi_{\lambda_{2}}^{\dagger}\left(\frac{2 k_{j}}{k^{+}}-\alpha_{j} \frac{\mathbf{\alpha} \cdot \mathbf{p}+\beta m}{p^{+}}-\frac{\mathbf{\alpha} \cdot(\mathbf{k}-\mathbf{p})-\beta m}{k^{+}-p^{+}} \alpha_{j}\right) \chi_{\lambda_{1}}\nonumber\\	
	 &&\times~\chi_{\lambda_{1}}^{\dagger}\left(\frac{2 k_{i}}{k^{+}}-\frac{\mathbf{\alpha} \cdot \mathbf{p}+\beta m}{p^{+}} \alpha_{i}-\alpha_{i} \frac{\mathbf{\alpha} \cdot(\mathbf{k}-\mathbf{p})-\beta {m}}{k^{+}-p^{+}}\right) \chi_{\lambda_{2}}
\end{eqnarray}
can be converted  (after summing over $\lambda_{1}$ and $\lambda_{2}$) to

\begin{eqnarray}
		\mathrm{tr}\Bigg[&\Lambda_{+}&\left(\frac{2 k_{j}}{k^{+}}-\alpha_{j} \frac{\mathbf{\alpha} \cdot \mathbf{p}+\beta m}{p^{+}}-\frac{\mathbf{\alpha} \cdot(\mathbf{k}-\mathbf{p})-\beta m}{k^{+}-p^{+}} \alpha_{j}\right)\nonumber \\ &\Lambda_{+} & \left(\frac{2 k_{i}}{k^{+}}-\frac{\mathbf{\alpha} \cdot \mathbf{p}+\beta m}{p^{+}} \alpha_{i}-\alpha_{i} \frac{\mathbf{\alpha} \cdot(\mathbf{k}-\mathbf{p})-\beta m}{k^{+}-p^{+}}\right)\Bigg]
			\label{eq:tr-g1}
\end{eqnarray} 
where
\begin{equation}
\sum_\lambda \chi_\lambda \chi_\lambda^\dagger = \Lambda_{+}
\end{equation}
has been used. After some Dirac algebra, the trace expression ~(\ref{eq:tr-g1}) becomes
\begin{equation}
	\label{eq:tr-g1-final}
	\frac{2}{k^{+2} \xi^{2}(1-\xi)^{2}}\left[-4 \xi(1-\xi) \tilde{p}_{i} \tilde{p}_{j}+\left({\tilde{\mathbf{p}}}^{2}+m^{2}\right) \delta_{i j}\right]
\end{equation} 
where $\xi\equiv p^+/k^+$, and $\tilde{\mathbf{p}}\equiv \mathbf{p}-\xi \mathbf{k}$. Substituting
 ~(\ref{eq:tr-g1-final}) back into Eq.~(\ref{eq:psig-1-Calc-1}) (and  changing  variables to 
 $\xi$ and $\tilde{\mathbf{p}}$),
\begin{eqnarray}
	|\psi_g^{1^\prime}\rangle&=&-\int_{\Lambda}^{e^{\delta Y}\Lambda} d k^{+} \int \frac{d^{2} \mathbf{k}}{ (2\pi)^{2}} \frac{d^{2} \tilde{\mathbf{p}}}{(2 \pi)^{2}} \int_{0}^{1} d \xi \nonumber \\
	&&\times\frac{g^{3} \rho^{a}(-\mathbf{k}) k_{i}\left(-4 \xi(1-\xi) \tilde{p}_{i} \tilde{p}_{j}+\left({\tilde{\mathbf{p}}}^{2}+m^{2}\right) \delta_{i j}\right)}{2 \sqrt{\pi k^{+}} \mathbf{k}^{4}\left(\xi(1-\xi) \mathbf{k}^{2}+\tilde{\mathbf{p}}^{2}+m^{2}\right) \xi(1-\xi)}\left|g_{j}^{a}(k)\right\rangle, 
\end{eqnarray}
where $\mathrm{tr}[t^a t^b]=\delta^{ab}/2$ is used. The
$d^2\tilde{\mathbf{p}}$ integral  is UV divergent. To regularize it, 
we use dimensional regularization with the  dimension $2 \to d=2-\epsilon$:
\begin{eqnarray}
	|\psi_g^{1^\prime}\rangle&=&-\mu^{\epsilon}\int_{\Lambda}^{e^{\delta Y}\Lambda} d k^{+} \int_{0}^{1} d \xi \int \frac{d^{2} \mathbf{k}}{ (2\pi)^{2}} \frac{d^{d} \tilde{\mathbf{p}}}{(2 \pi)^{d}} \nonumber \\
	&& \times \frac{g^{3} \rho^{a}(-\mathbf{k}) k_{i}\left(-4 \xi(1-\xi) \tilde{p}_{i} \tilde{p}_{j}+\left({\tilde{\mathbf{p}}}^{2}+m^{2}\right) \delta_{i j}\right)}{2 \sqrt{\pi k^{+}} \mathbf{k}^{4}\left(\xi(1-\xi) \mathbf{k}^{2}+\tilde{\mathbf{p}}^{2}+m^{2}\right) \xi(1-\xi)}\left|g_{j}^{a}(k)\right\rangle \nonumber\\
	&=&-\mu^{\epsilon}\int_{\Lambda}^{e^{\delta Y}\Lambda} d k^{+} \int_{0}^{1} d \xi \int \frac{d^{2} \mathbf{k}}{ (2\pi)^{2}} \frac{d^{d} \tilde{\mathbf{p}}}{(2 \pi)^{d}} \nonumber \\
	&& \times  \frac{g^{3} \rho^{a}(-\mathbf{k}) k_{i}\left((4\xi^2-4\xi+d){\tilde{\mathbf{p}}}^{2}+m^{2}d \right)}{2 \sqrt{\pi k^{+}} \mathbf{k}^{4}\left(\xi(1-\xi) \mathbf{k}^{2}+\tilde{\mathbf{p}}^{2}+m^{2}\right) \xi(1-\xi)d}\left|g_{i}^{a}(k)\right\rangle \nonumber\\
	&=&-\int_{\Lambda}^{e^{\delta Y}\Lambda} d k^{+} \int_{0}^{1} d \xi \int \frac{d^{2} \mathbf{k}}{ (2\pi)^{2}} \frac{1}{64 \pi^{7 / 2}} \frac{g^{3} \rho^{a}(-\mathbf{k}) k_{i}}{\sqrt{k^{+}} \mathbf{k}^{2}}\nonumber\\
	&&\times~\left[\left(4 \xi^{2}-4 \xi+2-\frac{4 m^{2}}{\mathbf{k}^{2}}\right)\left(-\frac{2}{\epsilon}+\ln \frac{\xi(1-\xi) \mathbf{k}^{2}+m^{2}}{{\mu_{\overline{MS}}}^{2}}-1\right)+2\right] \left|g_{i}^{a}(k)\right\rangle. \qquad
	\label{eq:psig1prime}
\end{eqnarray}
Now we can Fourier transform $\rho^a(-\mathbf{k})$ and $|g_i^a(k)\rangle$ (employing the conventions ~(\ref{eq:rho-fourier-conv}) and (\ref{eq:convention-state-k-to-x}))
\begin{eqnarray}
\left| \psi_g^{1^\prime}\right\rangle&=&\int_{\mathbf{xz}}\int_{\Lambda}^{e^{\delta Y}\Lambda}dk^+ \int_0^1 d\xi \int d^2\mathbf{k}\frac{e^{-i \mathbf{k}\cdot(\mathbf{x}-\mathbf{z})}}{128 \pi^{9 / 2}} \frac{g^{3} \rho^{a}(\mathbf{x}) k_{i}}{\sqrt{k^{+}} \mathbf{k}^{2}}\nonumber \\
&&\times~\left[\left(4 \xi^{2}-4 \xi+2-\frac{4 m^{2}}{\mathbf{k}^{2}}\right)\left(-\frac{2}{\epsilon}+\ln \frac{\xi(1-\xi) \mathbf{k}^{2}+m^{2}}{{\mu_{\overline{MS}}}^{2}}-1\right)+2\right] \left| g_i^a(k^+,\mathbf{z}) \right\rangle .\nonumber\\
\end{eqnarray}
In contrast to the massless case, however, the $\mathbf{k}$ integral is IR divergent.  
In order to regularize the divergence,  a "gluon mass"  $\delta$ can be introduced, i.e., we
make the following replacement
\begin{equation}
\frac{1}{\mathbf{k}^2} \to \frac{1}{\mathbf{k}^2+\delta^2}.
\end{equation}
With this IR regularization, $d^2 \mathbf{k}$ and $d\xi$ integrals can be performed (using the Fourier integral Eq.~(\ref{eq:int-gluon-mass-reg})), and we obtain the final expression in Eq.~(\ref{eq:psig-1-final}). As discussed in the main text, the IR regulator is eventually removed
by proper renormalization, which preserves  masslessness of the gluon.
 
\section{Subtractions} 
\label{app:subtractions}
This Appendix fills in some  gaps in obtaining Eq.~(\ref{eq:subtraction-1}) and (\ref{eq:subtraction}).\footnote{As is also mentioned in the Footnote~\ref{fn:algebra}, the algebraic manipulations here are  inspired by those used in Ref.~\cite{Kovchegov:2006wf} where running coupling effects are investigated with massless quarks.} We start from the first equality in Eq.~(\ref{eq:subtraction-1}):
\begin{eqnarray}
	&& \int_{\mathbf{z'}} K_{q\bar{q}}(\mathbf{x,y,z,z'}) \nonumber \\
	&& 	=\frac{g^4}{(2\pi)^{10}} \int_{\mathbf{z}^{\prime}} \int_{0}^{1} d \xi \int d^2 \mathbf{k} d^{2} \mathbf{p} d^{2} \mathbf{u} d^{2}\mathbf{v} e^{-i \mathbf{v} \cdot {Z}+i \mathbf{u} \cdot {Y}^{\prime}+i \mathbf{p} \cdot {Z}-i \mathbf{k} \cdot {X}^{\prime}} \nonumber \\ 
	&& \quad \times \frac{1}{\left((1-\xi) \mathbf{v}^{2}+\xi(\mathbf{u}-\mathbf{v})^{2}+m^{2}\right)\left((1-\xi) \mathbf{p}^{2}+\xi(\mathbf{k}-\mathbf{p})^{2}+m^{2}\right)} \Bigg[\xi^{2}\nonumber \\
	&&\quad +\xi(1-2 \xi)\left(\frac{\mathbf{k} \cdot \mathbf{p}}{\mathbf{k}^{2}}+\frac{\mathbf{u} \cdot \mathbf{v}}{\mathbf{u}^{2}}\right)  +\frac{(1-2 \xi)^{2} \mathbf{u} \cdot \mathbf{v} \cdot \mathbf{k} \cdot \mathbf{p}+\mathbf{u} \cdot \mathbf{k} \mathbf{v} \cdot \mathbf{p}-\mathbf{u} \cdot \mathbf{p} \mathbf{v} \cdot \mathbf{k}+m^{2} \mathbf{u} \cdot \mathbf{k}}{\mathbf{u}^{2} \mathbf{k}^{2}} \Bigg] \nonumber \\	\label{eq:subtraction-ap-1}
\end{eqnarray}
The $\mathbf{z}'$-integral produces a delta function (noticing that $Z=\mathbf{z-z'}$, etc.) 
\begin{equation}
	\int d\mathbf{z}'  \quad \to \quad (2\pi)^2 \delta(\mathbf{v-u-p+k})
\end{equation}
and the phase of the Fourier transform becomes
\begin{equation}
	e^{-i \mathbf{v} \cdot Z+i \mathbf{u} \cdot Y^{\prime}+i \mathbf{p} \cdot Z-i \mathbf{k} \cdot X^{\prime}} \quad \to \quad e^{-i \mathbf{k}\cdot X + i \mathbf{u}\cdot Y}.
\end{equation}
After performing the $\mathbf{z}'$ and $\mathbf{v}$-integrals
\begin{eqnarray}
	&& \int_{\mathbf{z'}} K_{q\bar{q}}(\mathbf{x,y,z,z'}) \nonumber \\
	&& 	=\frac{g^4}{(2\pi)^{10}} \int_{0}^{1} d \xi \int d^2 \mathbf{k} d^{2} \mathbf{p} d^{2} \mathbf{u}  e^{-i \mathbf{k}\cdot X + i \mathbf{u}\cdot Y} \nonumber \\ 
	&& \times \frac{1}{\mathbf{u}^{2} \mathbf{k}^{2}} \frac{1}{\left((1-\xi)(\mathbf{p}-\mathbf{k}+\mathbf{u})^{2}+\xi(-\mathbf{p}+\mathbf{k})^{2}+m^{2}\right)\left((1-\xi) \mathbf{p}^{2}+\xi(\mathbf{k}-\mathbf{p})^{2}+m^{2}\right)}\nonumber \\
	&&\times \Big[ \xi^{2} \mathbf{k}^{2} \mathbf{u}^{2}+\xi(1-2 \xi)\left(\mathbf{u}^{2} \mathbf{k} \cdot \mathbf{p}+\mathbf{k}^{2} \mathbf{u} \cdot(\mathbf{p}-\mathbf{k}+\mathbf{u})\right)+(1-2 \xi)^{2} \mathbf{u} \cdot(\mathbf{p}-\mathbf{k}+\mathbf{u}) \mathbf{k} \cdot \mathbf{p} \nonumber \\
	&&+\mathbf{u} \cdot \mathbf{k}(\mathbf{p}-\mathbf{k}+\mathbf{u}) \cdot \mathbf{p}-\mathbf{u} \cdot \mathbf{p}(\mathbf{p}-\mathbf{k}+\mathbf{u}) \cdot \mathbf{k}+m^{2} \mathbf{u} \cdot \mathbf{k} \Big].
	\label{eq:subtraction-ap2}
\end{eqnarray}
Now we can carry out the $\mathbf{p}$-integral, noticing that the phase of the Fourier transform is independent of $\mathbf{p}$. 
To make the integration kernel more symmetric, we further shift the momentum integrals as
\begin{equation}
	\mathbf{p}\to \mathbf{-p+k}, \quad \mathbf{u} \to \mathbf{u+p}, \quad \mathbf{k} \to \mathbf{k+p}.
\end{equation}
Then, after some algebra, we obtain
\begin{eqnarray}
	&& \int_{\mathbf{z'}} K_{q\bar{q}}(\mathbf{x,y,z,z'}) \nonumber \\
	&& 	=\frac{g^4}{(2\pi)^{8}} \int_{0}^{1} d \xi \int d^2 \mathbf{k} d^{2} \mathbf{p} d^{2} \mathbf{u}  e^{-i \mathbf{k}\cdot X + i \mathbf{u}\cdot Y} \nonumber \\ 
	&& \quad \times \frac{1}{(\mathbf{u}+\mathbf{p})^{2}(\mathbf{k}+\mathbf{p})^{2}} \bigg[
	1+\frac{\mathbf{k} \cdot \mathbf{p}-m^{2}}{(1-\xi) \mathbf{k}^{2}+\xi \mathbf{p}^{2}+m^{2}}+\frac{\mathbf{u} \cdot \mathbf{p}-m^{2}}{(1-\xi) \mathbf{u}^{2}+\xi \mathbf{p}^{2}+m^{2}}  \nonumber\\
	&& \quad+ \frac{\mathbf{p}^{2} \mathbf{u} \cdot \mathbf{k}+m^{2}(\mathbf{u}+\mathbf{p}) \cdot(\mathbf{k}+\mathbf{p})-m^{2}(\mathbf{u}+\mathbf{k}) \cdot \mathbf{p}+m^{4}}{\left((1-\xi) \mathbf{k}^{2}+\xi \mathbf{p}^{2}+m^{2}\right)\left((1-\xi) \mathbf{u}^{2}+\xi \mathbf{p}^{2}+m^{2}\right)}\bigg]\nonumber \\
	&& = \frac{g^4}{(2\pi)^4} \int_0^1 d\xi \int \frac{d^2\mathbf{k}}{2\pi} \frac{d^2\mathbf{p}}{(2\pi)^2} \frac{d^2\mathbf{u}}{2\pi} e^{-i \mathbf{k}\cdot X + i \mathbf{u}\cdot Y} \frac{1}{\mathbf{u}^2\mathbf{k}^2}  \nonumber \\     
	&&  \quad \times \Bigg[1+\frac{\mathbf{k} \cdot \mathbf{p}-\left(\mathbf{p}^{2}+m^{2}\right)}{(1-\xi)(\mathbf{k}-\mathbf{p})^{2}+\xi \mathbf{p}^{2}+m^{2}} +\frac{\mathbf{u} \cdot \mathbf{p}-\left(\mathbf{p}^{2}+m^{2}\right)}{(1-\xi)(\mathbf{u}-\mathbf{p})^{2}+\xi \mathbf{p}^{2}+m^{2}} \nonumber \\ &&\quad+\frac{\left(\mathbf{p}^{2}+m^{2}\right)(\mathbf{u}-\mathbf{p}) \cdot(\mathbf{k}-\mathbf{p})+\left(\mathbf{p}^{2}+m^{2}\right) m^{2}}{\left((1-\xi)(\mathbf{k}-\mathbf{p})^{2}+\xi \mathbf{p}^{2}+m^{2}\right)\left((1-\xi)(\mathbf{k}-\mathbf{p})^{2}+\xi \mathbf{p}^{2}+m^{2}\right)} \Bigg] 
	\label{eq:subtraction-ap3}
\end{eqnarray}
where for the second equality, we shift $\mathbf{u \text{~and~} k}$ back again with $\mathbf{u\to u-p, \, \,k\to k-p}$.
 Eq.~(\ref{eq:subtraction-1}) is then follows.

When attempting to carry out the $\mathbf{p}$-integral in Eq.~(\ref{eq:subtraction-1}), the last piece turns out to be  the most complicated one. It can be done as follows. Using the identity
\begin{equation}
\ln \frac{b}{a}	= (b-a) \int_{0}^{1} d t \frac{1}{(1-t) a+t b}
\end{equation}
the $\xi$-integral of the last piece simplifies to
\begin{eqnarray}
	&&\int d\xi \frac{\left(\mathbf{p}^{2}+m^{2}\right)(\mathbf{u}-\mathbf{p}) \cdot(\mathbf{k}-\mathbf{p})+\left(\mathbf{p}^{2}+m^{2}\right) m^{2}}{\left((1-\xi)(\mathbf{k}-\mathbf{p})^{2}+\xi \mathbf{p}^{2}+m^{2}\right)\left((1-\xi)(\mathbf{k}-\mathbf{p})^{2}+\xi \mathbf{p}^{2}+m^{2}\right)} \nonumber \\
	&& = \left[\left(\mathbf{p}^{2}+m^{2}\right)(\mathbf{u}-\mathbf{p}) \cdot(\mathbf{k}-\mathbf{p})+\left(\mathbf{p}^{2}+m^{2}\right) m^{2}\right] \frac{\ln \frac{(\mathbf{k}-\mathbf{p})^{2}+m^{2}}{(\mathbf{u}-\mathbf{p})^{2}+m^{2}}}{\left[(\mathbf{k}-\mathbf{p})^{2}-(\mathbf{u}-\mathbf{p})^{2}\right]\left(\mathbf{p}^{2}+m^{2}\right)} \nonumber \\
	&& =\frac{(\mathbf{k}-\mathbf{p}) \cdot(\mathbf{u}-\mathbf{p})+m^{2}}{(\mathbf{k}-\mathbf{p})^{2}-(\mathbf{u}-\mathbf{p})^{2}} \ln \frac{(\mathbf{k}-\mathbf{p})^{2}+m^{2}}{(\mathbf{u}-\mathbf{p})^{2}+m^{2}} \nonumber \\
	&&  =\int_{0}^{1} d \alpha \frac{(\mathbf{k}-\mathbf{p}) \cdot(\mathbf{u}-\mathbf{p})+m^{2}}{(1-\alpha)(\mathbf{u}-\mathbf{p})^{2}+\alpha(\mathbf{k}-\mathbf{p})^{2}+m^{2}}
\end{eqnarray}
where now the $\mathbf{p}$-integral of the integrand can be computed easily, yielding ($d = 2-\epsilon$)
\begin{eqnarray}
	&&\int \frac{d^d\mathbf{p}}{(2\pi)^d} \frac{(\mathbf{k}-\mathbf{p}) \cdot(\mathbf{u}-\mathbf{p})+m^{2}}{(1-\alpha)(\mathbf{u}-\mathbf{p})^{2}+\alpha(\mathbf{k}-\mathbf{p})^{2}+m^{2}}	\nonumber \\
	&&=-\frac{2}{4 \pi} \alpha(1-\alpha)(\mathbf{u}-\mathbf{k})^{2}\left(\frac{2}{\epsilon}-\ln \frac{\alpha(1-\alpha)(\mathbf{u}-\mathbf{k})^{2}+m^{2}}{\bar{\mu}^{2}}\right).
\end{eqnarray}

The $\mathbf{p}$-integrals for the remaining parts in Eq.~(\ref{eq:subtraction-1}) are straightforward (again by changing the integration variable from $\xi$ to $\alpha$). The final result is expressed in the first equality of ~(\ref{eq:subtraction}). The $\frac{1}{\epsilon}$ term from the dimensional regularization reflects the UV divergence of the integral, 
it is removed by the standard renormalization procedure. Finally, the second equality in Eq.~(\ref{eq:subtraction}) is obtained with the help of the double Fourier integrals (see Appendix \ref{app:integrals}).

\section{Integrals}
\label{app:integrals}
In addition to some integrals tabulated in  Ref.~\cite{Lublinsky:2016meo}, 
we also used the following integrals.

\subsection{$\xi$ Integrals}
\begin{eqnarray}
	f(x)&=&\frac{2}{3}+4x+\int_0^1 \left(4 \xi ^2-4 \xi -4 x+2\right) \log ((1-\xi ) \xi +x)d\xi \nonumber\\
	&=&\frac{2}{9} \left(\frac{12 \left(-8 x^2+2 x+1\right) \coth ^{-1}\left(\sqrt{4 x+1}\right)}{\sqrt{4 x+1}}+42 x+(6-18 x) \log (x)-10\right)\nonumber \\
\end{eqnarray}
with $f(0)=-\frac{20}{9}.$

\subsection{Fourier Integrals}
\begin{equation}
	\int \frac{d^{2} \mathbf{k}}{2\pi} \frac{d^{2} \mathbf{p}}{2\pi} \frac{1}{a \mathbf{k}^{2}+\mathbf{p}^{2}+m^{2}} e^{i \mathbf{k} \cdot \mathbf{x}+i \mathbf{p} \cdot \mathbf{y}} = \frac{1}{\mathbf{x}^{2}+a \mathbf{y}^{2}} m \sqrt{\left(\frac{\mathbf{x}^{2}}{a}+\mathbf{y}^{2}\right)} K_{1}\Bigg[m\sqrt{\left(\frac{\mathbf{x}^{2}}{a}+\mathbf{y}^{2}\right)}\Bigg],
\end{equation}

\begin{equation}
\int \frac{d^{2} \mathbf{k}}{2 \pi} \frac{d^{2} \mathbf{p}}{2 \pi} \frac{1}{a \mathbf{k}^{2}+\mathbf{p}^{2}+m^{2}} \frac{k_{i}}{\mathbf{k}^{2}} e^{i \mathbf{k} \cdot \mathbf{x}+i \mathbf{p} \cdot \mathbf{y}}=-i \frac{x^{i}}{\mathbf{x}^{2}} \left(K_{0}\left[m \sqrt{\frac{{\mathbf{x}^{2}+a \mathbf{y}^{2}}}{a}}\right]-K_{0}[m|\mathbf{y}|]\right),
\end{equation}

\begin{eqnarray}
	&&\int d^{2} \mathbf{k} d^{2} \mathbf{p} \frac{k^{i} p^{j}}{\mathbf{k}^{2}\left(a \mathbf{k}^{2}+\mathbf{p}^{2}+m^{2}\right)} e^{i \mathbf{k} \cdot \mathbf{x}+i \mathbf{p} \cdot \mathbf{y}}  =-4 \pi^{2} \frac{x^{i} y^j}{\mathbf{y}^{2}\left(a \mathbf{y}^{2}+\mathbf{x}^{2}\right)} \\
	&& \quad\times \Bigg( -\frac{a \mathbf{y}^{2}}{\mathbf{x}^{2}} m \sqrt{\frac{a \mathbf{y}^{2}+\mathbf{x}^{2}}{a}} K_{1}\left[m \sqrt{\frac{a{\mathbf{y}^{2}+\mathbf{x}^{2}}}{a}}\right]  +\frac{a \mathbf{y}^{2}+\mathbf{x}^{2}}{\mathbf{x}^2} m \mathbf{y} K_{1}[m|\mathbf{y}|] \Bigg),
	\end{eqnarray}

\begin{equation}
	\int \frac{d^{2} k}{2 \pi} \frac{k_{i}}{\mathbf{k}^{2}} \ln \left[\frac{\mathbf{k}^{2}+m^{2}}{\mu^{2}}\right] e^{i \mathbf{k} \cdot \mathbf{x}}=\frac{i x_{i}}{\mathbf{x}^{2}}\left[2 K_{0}\left[m |\mathbf{x}|\right]+\ln \frac{m^{2}}{\mu^{2}}\right],
	\label{eq:Fourier-ki-log}
\end{equation}

\begin{equation}
	\int \frac{d^{2} \mathbf{k}}{2 \pi} \ln \left(\mathbf{k}^{2}+m^{2}\right) e^{i \mathbf{k} \cdot \mathbf{x}}=-\frac{2 m}{|\mathbf{x}|} k_{1}[m|{\mathbf{x}}|],
\end{equation}

\begin{equation}
	\int \frac{d^{2} \mathbf{k}}{2 \pi} \frac{k_{i}}{\left(\mathbf{k}^{2}+\delta^2\right)^{2}} e^{i \mathbf{k} \cdot \mathbf{x}}=\frac{ix_i}{2}K_0[\delta|\mathbf{x}|],
	\label{eq:int-gluon-mass-reg-0}
\end{equation}

\begin{eqnarray}
	&&\int \frac{d^{2} \mathbf{k}}{2 \pi} \frac{k_{i}}{\left(\mathbf{k}^{2}+\delta^2\right)^{2}} \ln \left[\frac{\mathbf{k}^{2}+m^{2}}{\mu^{2}}\right] e^{i \mathbf{k} \cdot \mathbf{x}}\nonumber \\
	&&=\frac{i x_{i}}{2} K_{0}[\delta |\mathbf{x}|) \ln \left(\frac{m^{2}-\delta^2}{\mu^{2}}\right)+\frac{i x_{i}}{\mathbf{x}^{2}} {\delta} |\mathbf{x}| K_{1}[{\delta} |\mathbf{x}|] \frac{1}{m^{2}-\delta^2},
	\label{eq:int-gluon-mass-reg}
\end{eqnarray}

\begin{equation}
	\int \frac{d^{2} \mathbf{k}}{2 \pi} \frac{d^{2} \mathbf{p}}{2 \pi} \frac{1}{\mathbf{k}^{2}} \ln \frac{(\mathbf{k}-\mathbf{p})^{2}+m^{2}}{\mathbf{p}^{2}+m^{2}} e^{i \mathbf{k} \cdot \mathbf{x}+i \mathbf{p} \cdot \mathbf{y}} = \frac{1}{\mathbf{y}^{2}} \ln \left[\frac{(\mathbf{x}+\mathbf{y})^{2}}{\mathbf{x}^{2}}\right] m|\mathbf{y}| K_{1}[m|\mathbf{y}|],
	\label{eq:double-Fourier-one-over-ksquared}
\end{equation}

\begin{eqnarray}
	&&\int \frac{d^{2} \mathbf{k}}{2 \pi} \frac{d^{2}\mathbf{p}}{2 \pi} \frac{\mathbf{k \cdot p}}{\mathbf{k}^{2} \mathbf{p}^{2}} \ln \left[\frac{(\mathbf{k}-\mathbf{p})^{2}+m^{2}}{\mathbf{p}^{2}+m^{2}}\right] e^{i \mathbf{k} \cdot \mathbf{x}+i \mathbf{p} \cdot \mathbf{y}} \nonumber \\ 
&=& -{\rm Re}\left\{\int_{0}^{\mathbf{x}^{2}} d z^{2} \frac{1}{\left(x^{+} y^{-}\right)\left(x^{+} y^{-}+z^{2}\right)} m z K_{1}[m z]+\int_{\mathbf{y}^{2}}^{\infty} d z^{2} \frac{1}{z^{2}\left(x^{+} y^{-}+z^{2}\right)} m z K_{1}[m z]\right\} \nonumber\\
&&-\frac{2 \mathbf{x} \cdot \mathbf{y}}{\mathbf{x}^{2} \mathbf{y}^{2}}\left(K_{0}[m|\mathbf{x}|]-K_{0}[m|\mathbf{y}|]\right),
\label{eq:double-integrals-kdotp}
\end{eqnarray}
where $x^{\pm} \equiv |\mathbf{x}|e^{\pm i\eta}$ with $\eta$ being the polar angle of $x^1+ix^2$ on the complex plane (not to be confused with the light-cone coordinates), and the meaning of $y^{\pm}$ is similar. In particular, we have
\begin{equation}
	\mathrm{Re}[x^+y^-] = \mathbf{x} \cdot \mathbf{y}, \quad \frac{1}{x^+ y^-} = \frac{x^- y^+}{\mathbf{x}^2 \mathbf{y}^2}.
\end{equation}
Using  expansions of the Bessel functions ~(\ref{eq:bessel-taylor}), one can immediately obtain the massless limit  (can be also found in \cite{Fadin:2007de})
\begin{equation}
	\int \frac{d^{2} \mathbf{k}}{2 \pi} \frac{d^{2}\mathbf{p}}{2 \pi} \frac{\mathbf{k \cdot p}}{\mathbf{k}^{2} \mathbf{p}^{2}} \ln \left[\frac{(\mathbf{k}-\mathbf{p})^{2}}{\mathbf{p}^{2}}\right] e^{i \mathbf{k} \cdot \mathbf{x}+i \mathbf{p} \cdot \mathbf{y}} = -\frac{\mathbf{x \cdot y} }{\mathbf{x}^2 \mathbf{y}^2}\ln\left[\frac{(\mathbf{x}+\mathbf{y})^2}{\mathbf{x}^2} \right].
\end{equation}
Combining Eq.~(\ref{eq:Fourier-ki-log}) and (\ref{eq:double-integrals-kdotp}),
\begin{eqnarray}
	&&\int \frac{d^{2} \mathbf{k}}{2 \pi} \frac{d^{2}\mathbf{p}}{2 \pi} \frac{\mathbf{k \cdot p}}{\mathbf{k}^{2} \mathbf{p}^{2}} \ln \left[\frac{(\mathbf{k}-\mathbf{p})^{2}+m^{2}}{\mu^2}\right] e^{i \mathbf{k} \cdot \mathbf{x}+i \mathbf{p} \cdot \mathbf{y}} \nonumber \\ 
	&=& -{\rm Re}\left\{\int_{0}^{\mathbf{x}^{2}} d z^{2} \frac{1}{\left(x^{+} y^{-}\right)\left(x^{+} y^{-}+z^{2}\right)} m z K_{1}[m z]+\int_{\mathbf{y}^{2}}^{\infty} d z^{2} \frac{1}{z^{2}\left(x^{+} y^{-}+z^{2}\right)} m z K_{1}[m z]\right\} \nonumber\\
	&&-\frac{\mathbf{x} \cdot \mathbf{y}}{\mathbf{x}^{2} \mathbf{y}^{2}}\left(2K_{0}[m|\mathbf{x}|]+\ln\frac{m^2}{\mu^2}\right) \nonumber \\
	&& = -F(m,\mathbf{x},\mathbf{y}) -\frac{\mathbf{x} \cdot \mathbf{y}}{\mathbf{x}^{2} \mathbf{y}^{2}}\left(2K_{0}[m|\mathbf{x}|]+\ln\frac{m^2}{\mu^2}\right),
	\label{eq:double-integrals-kdotp-mu}
\end{eqnarray}
where
\begin{eqnarray}
	&&F(m,\mathbf{x},\mathbf{y}) \nonumber \\
	&&\equiv {\rm Re}\left\{\int_{0}^{\mathbf{x}^{2}} d z^{2} \frac{1}{\left(x^{+} y^{-}\right)\left(x^{+} y^{-}+z^{2}\right)} m z K_{1}[m z]+\int_{\mathbf{y}^{2}}^{\infty} d z^{2} \frac{1}{z^{2}\left(x^{+} y^{-}+z^{2}\right)} m z K_{1}[m z]\right\}. \nonumber \\
	{}
	\label{eq:F-def}
\end{eqnarray}
In the massless limit,
\begin{equation}
	F(m,\mathbf{x},\mathbf{y}) \Big|_{m\to 0} = \frac{\mathbf{x\cdot y}}{\mathbf{x}^2\mathbf{y}^2} \ln \left[\frac{(\mathbf{x}+\mathbf{y})^2}{\mathbf{y}^2}\right].
	\label{eq:F-massless-limit}
\end{equation}

The following is a brief sketch of how  Eq.~(\ref{eq:double-integrals-kdotp}) is derived. First, the Fourier transform of the delta function is used to decouple the cross term $(\mathbf{k}-\mathbf{p})^2$ between $\mathbf{k}$ and $\mathbf{p}$:
\begin{eqnarray}
	&&\int \frac{d^{2} \mathbf{k}}{2 \pi} \frac{d^{2}\mathbf{p}}{2 \pi} \frac{\mathbf{k \cdot p}}{\mathbf{k}^{2} \mathbf{p}^{2}} \ln \left[\frac{(\mathbf{k}-\mathbf{p})^{2}+m^{2}}{\mathbf{p}^{2}+m^{2}}\right] e^{i \mathbf{k} \cdot \mathbf{x}+i \mathbf{p} \cdot \mathbf{y}} \nonumber \\
	&& = \int\prod_{l=1}^{4} d^{2} \mathbf{z}_{l} \frac{d^{2} \mathbf{q}_{l}}{(2 \pi)^{2}} \int \frac{d^{2} \mathbf{k}}{2 \pi} \frac{d^{2} \mathbf{p}}{2 \pi}  \nonumber \\ 
	 && \quad \times \frac{\mathbf{q}_1 \cdot \mathbf{q}_2}{\mathbf{q}_1^2 \mathbf{q}_2^2}  \ln \frac{\mathbf{q}_{3}^{2}+m^{2}}{\mathbf{q}_{2}^{2}+m^{2}} e^{i\left(\mathbf{q}_{1}+\mathbf{k}\right) \cdot \mathbf{z}_{1}+i\left(\mathbf{q}_{2}+\mathbf{p}\right) \cdot \mathbf{z}_{2}+i\left(\mathbf{q}_{3}+\mathbf{k}-\mathbf{p}\right) \cdot \mathbf{z}_{3}+i\left(\mathbf{q}_{4}+\mathbf{p}\right) \cdot \mathbf{z}_{4}+i \mathbf{k} \cdot \mathbf{x}+i \mathbf{p} \cdot \mathbf{y}} \nonumber \\
	&& = \left(\int\prod_{l=1}^{4} d^{2} z_{l}\right) \frac{1}{(2 \pi)^{2}} \delta\left(\mathbf{z}_{1}+\mathbf{z}_{3}+\mathbf{x}\right) \delta\left(\mathbf{z}_{2}-\mathbf{z}_{3}+\mathbf{z}_{4}+\mathbf{y}\right) \frac{i \mathbf{z}_1\cdot i\mathbf{z}_2}{\mathbf{z}_{1}^{2} \mathbf{z}_{1}^{2}} \nonumber \\
	&& \quad \times  \left[ -\frac{2 m}{\left|\mathbf{z}_{3}\right|} K_{1}\left[m\left|\mathbf{z}_{3}\right|\right] \cdot 2 \pi \delta\left(\mathbf{z}_{4}\right) +\frac{2 m}{\left|\mathbf{z}_{4}\right|} K_{1}\left[m\left|\mathbf{z}_{4}\right|\right] \cdot 2 \pi \delta\left(\mathbf{z}_{3}\right) \right]     \nonumber \\
	&& = -\int \frac{d^{2} \mathbf{z}}{\pi} \frac{(\mathbf{z}-\mathbf{x})\cdot (\mathbf{z}+\mathbf{y})}{(\mathbf{z}-\mathbf{x})^{2} (\mathbf{z}+\mathbf{y})^{2}}  \frac{m}{|\mathbf{z}|} K_{1}[m|\mathbf{z}|]+\frac{x_{i}}{\mathbf{x}^{2}} \int \frac{d^{2} \mathbf{z}}{\pi} \frac{(z-y)_{i}}{(\mathbf{z}-\mathbf{y})^{2}} \frac{m}{|\mathbf{z}|} K_{1}(m|\mathbf{z}|) 
	\label{eq:kdotp-derivation-1}
\end{eqnarray}
The angular part of the integral in the last line of ~(\ref{eq:kdotp-derivation-1}) can  be now computed (for more details of the methods, please refer  to ~\cite{Fadin:2012my}). Decomposing the $\mathbf{z}$-integral into radial and angular parts
\begin{equation}
	\int \frac{d^2\mathbf{z}}{\pi} (\cdots) = \int_0^\infty dz^2 \int_{-\pi}^{\pi} \frac{d \phi}{2\pi} (\cdots),
\end{equation}
the angular integrals in Eq.~(\ref{eq:kdotp-derivation-1}) are 
\begin{eqnarray}
	\int_{-\pi}^{\pi} \frac{d \phi}{2 \pi} \frac{(\mathbf{z}-\mathbf{x})\cdot (\mathbf{z}-\mathbf{x})}{(\mathbf{z}+\mathbf{y})^{2} (\mathbf{z}+\mathbf{y})^{2}} &= & \left(\theta\left(\mathbf{x}^{2}-z^{2}\right)-\theta\left(z^{2}-\mathbf{y}^{2}\right)\right)\left(\frac{1}{-x^{+} y^{-}-z^{2}}+\frac{1}{-x^{-} y^{+}-z^{2}}\right), \nonumber \\
\end{eqnarray}
and
\begin{equation}
 \int \frac{d \phi}{2 \pi} \frac{\mathbf{z}-\mathbf{x}}{(\mathbf{z}-\mathbf{x})^{2}}=-\frac{\mathbf{x}}{\mathbf{x}^{2}} \theta\left(\mathbf{x}^{2}-\mathbf{z}^{2}\right),
\end{equation}
where $\theta$ is the Heaviside theta function. With the angular integral done, the derivation of Eq.~(\ref{eq:double-integrals-kdotp}) from Eq.~(\ref{eq:kdotp-derivation-1}) is straightforward. Notice that there is an exact UV divergence cancellation (as $\mathbf{z} \to \mathbf{0}$) in the last line of Eq.~(\ref{eq:kdotp-derivation-1}). 



\end{document}